\documentstyle{article}

\newcommand{\be}{\begin{equation}}
\newcommand{\ee}{\end{equation}}
\newcommand{\bea}{\begin{eqnarray}}
\newcommand{\eea}{\end{eqnarray}}

\begin{document}

\title{On estimation of the post-Newtonian parameters in
the gravitational-wave emission of a coalescing binary}

\author{Andrzej Kr\'olak
\\{Max-Planck-Research-Group Gravitational Theory}
\\{at the Friedrich-Schiller-University, 07743 Jena, Germany}
\thanks{Permanent address: Institute of Mathematics,
Polish Academy of Sciences, \'Sniadeckich 8, 00-950 Warsaw, Poland.}
\and Kostas D. Kokkotas
\\{Section Astrophysics, Astronomy and Mechanics,}
\\{Department of Physics, Aristotle University of Thessaloniki,}
\\{540 06 Thessaloniki, Macedonia, Greece}
\and Gerhard Sch\"afer
\\{Max-Planck-Research-Group Gravitational Theory}
\\{at the Friedrich-Schiller-University, 07743 Jena, Germany}}

\date{}
\maketitle
\newpage

\begin{abstract}
The effect of the recently obtained 2nd post-Newtonian corrections
on the accuracy of estimation of parameters of the
gravitational-wave signal from a coalescing binary is investigated.
It is shown that addition of this correction degrades considerably
the accuracy of determination of individual masses of the members
of the binary. However the chirp mass and the time parameter in
the signal is still determined to a very good accuracy.
The possibility of estimation of effects of other theories of gravity
is investigated.
The performance of the Newtonian filter is investigated and it is
compared with performance of post-Newtonian search templates introduced
recently. It is shown that both search templates can extract accurately
useful information about the binary.

PACS numbers: 04.30.+x,04.80.+x,95.85.5z,97.80.Af
\end{abstract}

\newpage

\section{Introduction}

It is currently believed that the gravitational waves
that come from the final
stages of the evolution of compact binaries just before their coalescence
are very likely signals to be detected by long-arm laser interferometers
\cite{T}.
The reason is that in the case of binary systems we can predict the
gravitational waveform very well; and the amplitudes are reasonably high
for sources at distances out to 200 Mpc. An estimate based on the number of
compact binaries known in our galaxy and extrapolated to the rest of
the Universe
shows that there should be one neutron star compact binary coalescence per year
out to the distance of 200Mpc \cite{N,Ph}. This estimate is a safe lower
bound on
the rate of binary coalescence. Arguments based on progenitor
evolution scenarios suggest
that there should be 100 of two neutron star coalescences,
5 neutron star - black
hole coalescences, and 0.5 two black hole coalescences out to 200Mpc
\cite{TY}.
The waveform derived using the
quadrupole formula has been known for quite some time \cite{P}.
A standard optimal method to detect the signal from a coalescing binary
in a noisy data set and to estimate its parameters is to correlate
the data with the filter matched to the signal and vary the parameters of the
filter until the correlation is maximal. The parameters of the filter
that maximize the correlation are estimators for the parameters of the signal.
The detailed algorithms and the performance of the matched-filtering
method in application to coalescing binary gravitational-wave signal
has been investigated by several authors, e.g. \cite{FC,KLM1,SD,KKT,CF}.
It has recently been realized \cite{Cal} that the correlation is very
sensitive even to very small variations of the phase of the filter because
of the large number of cycles in the signal.
Consequently the addition of small corrections to the phase of the signal due
to the post-Newtonian effects decreases the correlation considerably.
Thus the post-Newtonian effects in the coalescing binary waveform can
be detected and estimated to a much higher accuracy than it was thought
before \cite{Ks}. This opens up new prospects but also considerable
data analysis challenges for the LIGO, VIRGO, and GEO600 projects
which are rapidly progressing.
It was also found \cite{Cal} that the post-Newtonian series
is not converging rapidly for a binary near coalescence. Hence
higher post-Newtonian corrections will affect the correlation.
Currently three post-Newtonian corrections to the quadrupole formula are
already known \cite{BDIWW} and the calculation of further ones is
in progress.
In this article we analyse the estimation of parameters of the 2nd
post-Newtonian signal. This part of work complements
a recent detailed analysis of the 3/2 post-Newtonian signal
performed recently in Ref.\cite{CF}.  We also examine the
detectability of the post-Newtonian signal and estimation of its parameters
using the Newtonian waveform as a filter. This filter can be used
as the simplest search template. We compare the Newtonian search templates with
the post-Newtonian search templates recently investigated in Ref.\cite{Apo}.

The paper is organized as follows. In the first part of Section 2
we present the gravitational
wave signal from a binary system to the currently known 2nd post-Newtonian
order. In this work we analyse the signal in the ``restricted"
post-Newtonian approximation (i.e. only the phase of the signal is
given to the 2nd post-Newtonian accuracy whereas the amplitude of the signal
is calculated from the quadrupole formula),
we assume circularized orbits, and we assume the
spin parameters to be constant. In the second part we briefly describe
the optimal method of detection
of such a signal in noise and the maximum likelihood (ML) method to estimate
the parameters of the signal. We derive a number of properties of the
ML estimators of the parameters of our signal and we
examine the bounds on their variances. Our analysis is based on the Cramer-Rao
bound. In the third part we give the approximate rms errors of the estimators
for the signal at various post-Newtonian orders. In the fourth part of
Section 2 we consider the effects of other theories of gravity and
their detectability from gravitational-wave measurements. We consider
Jordan-Fiertz-Brans-Dicke theory and Damour-Esposito-Far\'ese biscalar
tensor theory. In Section 3 we consider the so called ``search templates"
introduced in Ref.\cite{Cal}.
These are simple filters containing as few parameters as possible to
effectively detect the multi-parameter signals.
In the first part of Section 3
we analyse the simplest search template - the Newtonian filter which is the
waveform of the gravitational signal from a binary in the quadrupole
approximation. We examine the Newtonian filter as a tool both to detect
the signal and also to determine its nature. In the second part of Section 3
we compare the Newtonian filter with the other search template analysed
recently \cite{Apo} based on the full post-Newtonian signal.
In Section 4 we summarize conclusions from our results.
A number of results is left to appendices. In Appendix A we examine
the first order effects on the phase of the signal due to eccentricity.
In Appendix B we give numerical values of the covariance matrices at
various post-Newtonian orders. In Appendix C we give certain detailed
formulae for the Damour-Esposito-Far\'ese theory.
In Appendix D we briefly review
the theory of optimal detection of known signal in noise and we generalize it
to non-optimal detection. In Appendix E we give a useful analytic
approximation to the correlation integral of the optimal filter with the
signal from a binary.

The units are chosen such that $G = c = 1$.

\section{Post-Newtonian effects}
\subsection{Gravitational wave signal from a coalescing binary}

Let us first give the formula for the gravitational waveform of a binary
with the three currently known post-Newtonian corrections.
We make the following approximations.
We work within the so called ``restricted" post-Newtonian
approximation i.e. we only include the post-Newtonian
corrections to the phase of the signal keeping the amplitude in its
Newtonian form; this is because the effect of the phase on
the correlation is dominant.
The inclusion of post-Newtonian effects in amplitudes
will not qualitatively change our results.
Due to the effect of rapid circularization of the orbit by radiation reaction
one can assume that the orbit is quasicircular. For example in the case
of the gravitational wave signal from the Hulse-Taylor binary pulsar at the
characteristic frequency of the detector for such a signal of around 47Hz
the eccentricity $e$ would be
$ \sim 10^{-6}$. Moreover the first order contribution to the phase of the
signal due to
eccentricity goes like $e^2$.
Nevertheless for completeness we include the first order correction
to the signal due to eccentricity in our formulae. We give a detailed
derivation
of this correction in Appendix A.
We neglect the tidal effects. All tidal contributions to the gravitational
wave signal from a coalescing binary were estimated to be small \cite{Ks,LW}.
There is also a small additional contribution to the phase
due to tail effects which detectability
has been considered in detail \cite{BS1} and was found to be small.
This correction is formally of the 4th post-Newtonian order
and consequently we neglect it in the present analysis.

With these approximations the waveform, as a function of time, is given
by the following expression,
\begin{equation}
h(t) =  A \, f(t)^{2/3} \cos[2\pi \int^t_{t_a}f(t')dt' - \phi],
\end{equation}
where
\be
A = \frac{8}{5}\pi^{2/3}\frac{\mu m^{2/3}}{R}
\ee
and where $\phi$ is an arbitrary phase, $\mu$ and $m$ are the reduced
and the total mass of the binary, respectively;
$t_a$ is a time parameter and $R$ is the distance to the source.
$A$ is the rms average amplitude over all Euler angles determining the
position of the binary on the sky
and the inclination angle between the plane of the orbit
of the binary and the line of sight. The rms amplitude $A$ is
2/5 of the maximum possible amplitude.
The characteristic time for the evolution of the binary to the
currently known 2nd post-Newtonian order is given by
\begin{eqnarray}
& & \tau_{2PN} := \frac{f}{df/dt} =
\frac{5}{96}\frac{1}{\mu m^{2/3}}\frac{1}{(\pi f)^{8/3}} \times
\\ \label{ttau}
& & [1 - \frac{157}{24}\frac{I_e}{f^{19/9}} +
(\frac{743}{336} + \frac{11}{4}\frac{\mu}{m})(\pi m f)^{2/3} -
(4\pi - s_o)(\pi m f) + \\ \nonumber
& & (\frac{3058673}{1016064} + \frac{5429}{1008}\frac{\mu}{m}
 + \frac{617}{144}(\frac{\mu}{m})^2 + s_s\frac{m}{\mu})(\pi m f)^{4/3}]
\end{eqnarray}
where $I_e$ is the asymptotic eccentricity invariant,
\be
I_e = e_0^2 f_0^{19/9},
\ee
$e_0$ is the eccentricity of the binary at gravitational frequency $f_0$
(see Appendix A for derivation and explanations).
The quantities $s_o$ and $s_s$ are spin-orbit and
spin-spin parameters respectively.
They are given by the formula
\begin{eqnarray}
s_o & = & \frac{113}{12}(s_{1} + s_{2}) +
\frac{25}{4}(s_{1}\frac{m_2}{m_1} + s_{2}\frac{m_1}{m_2}) \\
s_s & = & \frac{247}{48}{\bf s_1}\cdot{\bf s_2} - \frac{721}{48} s_1 s_2 \\
{\bf s_1} & = & \frac{{\bf S_1}}{m^2},\hspace{3mm}
{\bf s_2} = \frac{{\bf S_2}}{m^2}\\
s_1 & = & {\bf L}\cdot{\bf s_1},\hspace{3mm} s_2 = {\bf L}\cdot{\bf s_2},
\end{eqnarray}
where
${\bf L}$ is the total orbital angular momentum and ${\bf S_1}$, ${\bf S_2}$
are the spin angular momenta of the two bodies.
The terms in square brackets in Eq.(\ref{ttau}) are respectively: at lowest
order, Newtonian (quadrupole); at order $f^{-19/9}$, lowest order
contribution due to eccentricity (see Appendix A); at order $f^{2/3}$,
1PN \cite{WW}; at order $f$, the non-linear effect of ''tails" of the wave
($4\pi$ term) \cite{P1,BD,W,BS}), and spin-orbit effects \cite{KWW}; and
at order  $f^{4/3}$, 2PN \cite{BDIWW} and spin-spin effects \cite{KWW}.

In general the spin parameters vary with time. It was shown
\cite{CF} that $s_o$ is nearly conserved, it never deviates from its average
value by more than $\sim$ 0.25. Moreover the time dependent part of
the spin parameter is oscillatory what reduces considerably its influence
on the phase of the signal \cite{CF}. In this work we shall assume
that both spin-orbit and spin-spin parameters are constant.
We also neglect the effect of the precession of the
orbital plane due to spin on the waveform. The effects of the spin on the
waveform of the signal from an inspiralling binary have been investigated
in detail in \cite{ACST}.
If we take the available estimate of the moment of inertia for the pulsar
in the Hulse-Taylor binary and assume masses of the neutron stars
in the binary of 1.4 solar
masses then $s_o \simeq 4.8\times 10^{-2}$ and $s_s \simeq 2.4\times 10^{-5}$.
If such values are typical then spin effects will make negligible
contributions to the phase of the signal. However this may not be the case
for binaries containing black holes. Moreover if cosmic censorship is
violated and black holes rotate at a higher rate than allowed by maximally
rotating Kerr black hole the spin effects will significantly
affect the gravitational waveform.

In the analysis of the detection of the above signal and estimation of its
parameters it is convenient to work in the Fourier domain.
The expression for the Fourier transform of our signal in the stationary phase
approximation is given by (cf.\cite{T,DKSW,FC,CF})
\begin{eqnarray}
& &\tilde{h} = \tilde{A} f^{-7/6} \exp i[2\pi f t_{a}
- \phi - \pi/4 + \\ \nonumber
& &\frac{5}{48}(a(f;f_a) k + a_e(f;f_a) k_e + a_1(f;f_a) k_1 +
a_{3/2}(f;f_a) k_{3/2} + a_2(f;f_a) k_2)],
\label{Fou}
\end{eqnarray}
for $f > 0$ and by the complex conjugate of the above expression
for $f < 0$ where
\begin{eqnarray}
& &\tilde{A} = \frac{1}{(30)^{1/2}}\frac{1}{\pi^{2/3}}
\frac{\mu^{1/2}m^{1/3}}{R},
\\
& &k = \frac{1}{\mu m^{2/3}} ,\\
& &k_e = \frac{1}{\mu m^{2/3}}e_0^2(\pi f_0)^{19/9}, \\
& &k_1 = \frac{1}{\mu}(\frac{743}{336} + \frac{11}{4}\nu),\\
& &k_{3/2} = \frac{m^{1/3}}{\mu}(4\pi - s_o), \\
& &k_2 = \frac{m^{4/3}}{\mu}\left(\frac{3058673}{1016064}
+ \frac{5429}{1008}\nu +
\frac{617}{144}\nu^2 + \frac{s_s}{\nu}\right),\\
& &a(f;f_a) = \frac{9}{40}\frac{1}{(\pi f)^{5/3}} +
\frac{3}{8}\frac{\pi f}{(\pi f_a)^{8/3}}
- \frac{3}{5}\frac{1}{(\pi f_a)^{5/3}}, \\
& &a_e(f;f_a) = - \frac{157}{24}\left(\frac{81}{1462}\frac{1}
{(\pi f)^{34/9}} +
\frac{9}{43}\frac{\pi f}{(\pi f_a)^{43/9}} -
\frac{9}{34}\frac{1}{(\pi f_a)^{34/9}}\right), \\
& &a_1(f;f_a) = \frac{1}{2}\frac{1}{\pi f} +
\frac{1}{2}\frac{\pi f}{(\pi f_a)^2} - \frac{1}{\pi f_a}, \\
& &a_{3/2}(f;f_a) = - \left( \frac{9}{10}\frac{1}{(\pi f)^{2/3}} +
\frac{3}{5}\frac{\pi f}{(\pi f_a)^{5/3}}
- \frac{3}{2}\frac{1}{(\pi f_a)^{2/3}}\right), \\
& &a_2(f;f_a) = \frac{9}{4}\frac{1}{(\pi f)^{1/3}} +
\frac{3}{4}\frac{\pi f}{(\pi f_a)^{4/3}} - \frac{3}{(\pi f_a)^{1/3}}
\end{eqnarray}
hold and where $f_a = f(t_a)$.

The stationary phase approximation, Eq.(\ref{Fou}), is an excellent
approximation of the Fourier transform of the signal for frequencies
which are not influenced by the finite time window of the measurement.
In the above expressions for the gravitational wave signal
from a binary
we can make an arbitrary choice of the time parameter and the phase
of the signal.

We also point out that going from the time to the frequency domain
we have made yet another approximation. Namely we have taken
the modulus $\mid\tilde{h}\mid$ of the Fourier
transform to be the Newtonian one i.e. $\mid\tilde{h}\mid\sim f^{-7/6}$.
In the stationary phase approximation $\mid\tilde{h}\mid$  goes like
$1/\sqrt{\dot{f}}$
and consequently by Eq.(\ref{tau}) there would be other
powers of frequency due to the post-Newtonian effects.
We neglect those additional terms since the
post-Newtonian corrections to the phase have the dominant effect. The inclusion
of the post-Newtonian amplitudes to the signal will not qualitatively
change the results of
this work.

A convenient parameter is the {\em chirp mass}
defined as ${\cal M} := k^{-3/5}$.
In the quadrupole approximation the gravitational
wave signal from a binary is entirely determined by the chirp mass.

We shall consider three models of binaries: neutron star/neutron star
(NS-NS), neutron star/black hole (NS-BH), and black hole/black hole (BH-BH)
binaries with parameters summarized in Table I.

{\bf Table I.}
{\em Numerical values of the parameters of the three fiducial binary systems.
Black holes are of 10 solar masses and neutron stars are of 1.4 solar
masses. Spin parameters are assumed to be constant. Spin
for neutron stars was calculated from the typical estimate of the moment of
inertia $I$ for a neutron star of $I = 10^{38}$ kg m$^2$.}

\bigskip

{\bf A}
\begin{tabular}{|c|c|c|c|c|c|c|c|} \hline
Binary & $m_1 [M_{\odot}]$ & $m_2 [M_{\odot}]$ & ${\cal M} [M_{\odot}]$
& $s_1$
& $s_2$ & $s_o$ & $s_s$ \\ \hline
1. NS-NS  & 1.4 & 1.4 & 1.2 & 1.5$\times10^{-3}$ & 1.5$\times10^{-3}$ &
4.8$\times 10^{-2}$ & 2.4$\times 10^{-5}$ \\
2. NS-BH  & 1.4 & 10 & 3.0 & 3.0$\times 10^{-5}$ & 0.38 &
4.0 & 3.5 $\times 10^{-4}$\\
3. BH-BH  & 10 & 10 & 8.7 & 0.13 & 0.13 & 3.9 & 0.15 \\ \hline
\end{tabular}

\bigskip

\begin{center}
{\bf B}
\begin{tabular}{|c|c|c|c|c|} \hline
Binary & $k [M_{\odot}^{-5/3}]$ & $k_1 [M_{\odot}^{-1}]$ &
$k_{3/2} [M_{\odot}^{-2/3}]$ & $k_2 [M_{\odot}^{-1/3}]$ \\ \hline
1. NS-NS  & 0.72 & 4.1 & 25 & 13 \\
2. NS-BH  & 0.16 & 2.0 & 16 & 15 \\
3. BH-BH  & 2.7$\times10^{-2}$ & 0.58 & 4.7 & 7.7\\ \hline
\end{tabular}
\end{center}
\vspace{10mm}

where $M_{\odot}$ means solar mass.

For neutron stars we calculated the spin  using the available estimate
of the moment of inertia for the neutron star in the binary pulsar PSR1916+19.
We have taken black holes to be spinning at half the maximum rate
(i.e. $s_i = 0.5 m_i^2/m^2$).
The orbital momenta vectors were assumed to be parallel to the spin vectors.

To have an idea of the size of the post-Newtonian corrections
in the gravitational wave signal from a binary when it enters the observation
window of the laser interferometer we have evaluated
the characteristic time $\tau_{2PN}$
for the above three models at the frequency $f'_o = 47$ Hz
which is the characteristic
frequency of the detector for this signal (see below).
We have made explicit the contributions to the
characteristic time from the three post-Newtonian corrections.
\begin{eqnarray}
\tau^1_{2PN}=44 (1 + 0.046 \mbox{[from 1pn]}
- 0.025 \mbox{[from 3/2pn]} + 0.0012 \mbox{[from 2pn]}) \mbox{sec} \\
\tau^2_{2PN}=9.9 (1 + 0.10 \mbox{[from 1pn]}
- 0.071 \mbox{[from 3/2pn]} + 0.0060 \mbox{[from 2pn]}) \mbox{sec} \\
\tau^3_{2PN}=1.7 (1 + 0.17 \mbox{[from 1pn]}
- 0.12 \mbox{[from 3/2pn]} + 0.018 \mbox{[from 2pn]}) \mbox{sec}
\end{eqnarray}
One concludes from the above numbers that for the earth-based laser
interferometers
post-Newtonian corrections are significant. Moreover several things
are apparent.
The quadrupole term is dominant for all the three models. This indicates
a very good accuracy of the quadrupole formula even in the regime of strongly
gravitating bodies. This has been noticed in other studies
for example in the numerical investigation of the gravitational
wave emission from the two black hole collisions \cite{S}.
The difference in size between the 1st post-Newtonian correction and the
3/2 post-Newtonian correction (tail term) is rather small.
They differ by a factor of 2 for NS-NS binary and only by a factor of
around 1.5 for
binaries with a black hole. The second post-Newtonian correction is noticably
smaller then the 3/2 post-Newtonian correction.
The difference varies form a factor
of 20 for a NS-NS binary to a factor of 7 for a BH-BH binary.
The convergence of the post-Newtonian series appears to be worst for BH-BH
binaries and in this case it would be desirable to have accurate numerical
waveforms and not only the ones based on the post-Newtonian approximation.
Such waveforms should be available
as a result of the numerical projects such as
Grand Challenge project currently under way in the United States.

\subsection{Detection of the signal and estimation of its parameters}

For the purpose of this investigation we shall use a fit to the
total spectral density $S_h(f)$ of the noise in the advanced LIGO detectors,
devised in \cite{CF}. This fit comprises seismic, thermal, shot, and
quantum noises
in the detector.
\begin{equation}
S_h(f) = S_o((f_o/f)^4 + 2(1 + (f/f_o)^2))/5,
\label{Sh}
\end{equation}
where $f_o = 70$Hz and $S_o = 3\times 10^{-48}$Hz$^{-1}$.
It is an excellent approximation
to the detailed formulae for various noises given in \cite{FC}.
The {\em sensitivity function} $Sen(f)$ of the detector is defined
as $1/S_h(f)$.
The sensitivity function has the maximum at frequency $f_o$ given above
and its {\em half width half magnitude} (HWHM) $\sigma_o$ is around 48Hz.

To determine whether or not there is a signal in a noisy data set we use the
Neyman-Pearson test (see Appendix D).
When the noise in the detector is Gaussian the Neyman-Pearson test
is the {\em correlator test}. It consists of
linear filtering the data with the filter which Fourier transform
is the Fourier
transform of the signal divided by the spectral density of the noise \cite{He}.
The signal-to-noise ratio $d$ that can be achieved by optimal filtering
is given by $d = (h|h)^{1/2}$ where following \cite{CF} the scalar product
$(h_1|h_2)$ is defined by
\begin{equation}
(h1|h2)= 4 \Re\int^{\infty}_{f_i}\frac{\tilde{h}_1^{*}\tilde{h}_2}{S_h(f)} df,
\end{equation}
where $\Re$ denotes the real part.
Thus we have
\begin{equation}
d^2 = 4\tilde{A}^2\int^{\infty}_{f_i}\frac{df}{S_h(f) f^{7/3}}.
\end{equation}
We shall call the integrand  of the above signal-to-noise integral
{\em signal sensitivity function} and we denote it by $Ind(f)$.
This function  has the maximum
at the frequency $f_o'$ where $f_o' = 47$Hz and its HWHM
$\sigma_o'$ is $\simeq$ 26Hz, around
half of that of the sensitivity function.
This is the signal-to-noise ratio after filtering of the data.
We see that linear filtering introduces an effective narrowing
of the detector bandwidth \cite{DS}.

In the case of our chirp signal the linear filtering increases the
signal-to-noise ratio by an amount given roughly by the square root
of the number $n(f)$ of cycles spent near the frequency $f_o'$ where $n(f)$
is defined by \cite{T}
\begin{equation}
n(f) := f \tau \simeq \frac{5}{96\pi} \frac{1}{{\cal M}^{5/3}}
\frac{1}{(\pi f)^{5/3}}
\end{equation}
Consequently the effectiveness of matched filtering falls with the chirp mass.
On the other hand the amplitude $A$ of the signal increases with
the chirp mass like ${\cal M}^{5/3}$
and the overall factor in the signal-to-noise ratio
increases as ${\cal M}^{5/6}$.
This is born out by the amplitude $\tilde{A}$ of the Fourier transform.
Thus the probability of detection
of binaries with the same rate of occurrence increases with the chirp mass.

To estimate the parameters of the signal it is proposed to use the
{\em maximum likelihood
estimation} (MLE) \cite{D}. It is by no means guaranteed that this is the
best or the ultimate method.
It may sometimes fail to give an estimate and other methods may
lead to more accurate estimates.
The MLE method consists of maximizing the likelihood ratio
with respect to the parameters of the filter.
In the case of the Gaussian noise the logarithm of the likelihood ratio
$\Lambda$ is given by \cite{D}
\begin{equation}
\ln\Lambda = (x|h_F) - \frac{1}{2}(h_F|h_F)
\end{equation}
where $h_F$ which we call the filter has the form of the signal
but with arbitrary parameters
and $x$ are the data. We assume that the noise $n$ in the detector
is additive i.e. $x = h + n$.
The maximum likelihood (ML) estimators of the parameters of the signal are
given by the following set of differential equations providing
that one can differentiate
under the integration sign of the scalar product defined above.
\begin{equation}
(x - h_F|h_F,i) = 0,
\label{ml}
\end{equation}
where $h_F,i$ is the derivative of $h_F$ with respect to the $i$th parameter.
Rarely these equations can be solved analytically. It was shown \cite{KLM1}
that in the case of the signal from a binary within the stationary
phase approximation analytic expressions can be obtained
for the maximum likelihood estimators of the amplitude and the phase.

The ML estimators are random variables since they depend on the noise.
It is important to know
the statistical properties of these estimators and their probability
distributions so
that we can determine how well they estimate the true values of the parameters.
The most important quantities are the {\em expectation value}
of the estimator and its {\em variance}.
We would like to have the expectation value of the estimator to be as close
as possible to
the true value of the parameter and we would like the variance of the
estimator
to be as small as possible. The difference between the expectation value of
an estimator
of a parameter and the true value of the parameter is called the {\em bias}
of the estimator.
The ML estimator is not guaranteed to be either unbiased or minimum variance.
We have the following useful general inequality called
the Cramer-Rao inequality \cite{L}
that gives lower bound of the variance
of estimators. Let $(\theta_i)$ be a set of n parameters
and let $\theta_I$ be one of the parameters
then the variance of its estimator $\hat\theta_I$
satisfies the following inequality
\begin{equation}
Var[\hat\theta_I] \geq (\Gamma^{-1})_{ij}\alpha^i\alpha^j,
\end{equation}
where $\alpha^i$ and $\Gamma^{ij}$  are given by
\begin{eqnarray}
\alpha^i = \frac{\partial E[\hat{\theta_I}]}{\partial\theta_i}, \\
\Gamma^{ij} =
E[\frac{\partial\ln\Lambda}{\partial\theta_i}\frac{\partial\ln\Lambda}
{\partial\theta_j}],
\end{eqnarray}
where E is the expectation value.
The matrix ${\bf\Gamma}$ is called the Fisher information matrix and its
inverse is called the covariance matrix.
One easily sees from the above inequality that when an estimator
$\hat\theta_I$ is unbiased then
the lower bound on its variance is given by the (II) component of the
covariance matrix.
For this inequality to hold certain mathematical assumption
must be fulfilled \cite{L}.

\bigskip

1. The likelihood ratio must be a differentiable function
   with respect to all the parameters $\theta_i$.

2. The order of differentiation with respect to parameters and
   the integration in the expectation value integral must be interchangeable.

3. The variances of the estimators must be bounded.

4. The Fisher information matrix must be positive definite.

\bigskip

The Cramer-Rao inequality is very general.
It holds no matter what is the probability distribution
of the data and it applies to any estimator providing the regularity
conditions mentioned above
are fulfilled.
The above inequality guarantees only that the variance of an estimator
is greater then a certain amount. It is important for us to know
how well the right hand side of the Cramer-Rao
inequality approximates the actual variance of an estimator.
It was shown \cite{He,FC} that in the
case of Gaussian noise and in the limit of high signal to noise ratio
$d$ to the first order
the maximum likelihood estimators are Gaussian and moreover they are
unbiased and their
covariances are given by the covariance matrix defined above.
In statistical literature there also exists a series of refined
Cramer-Rao bounds called
Battacharyya bounds \cite{L}. However in our case a useful approach
to have an idea of the accuracy of the Cramer-Rao lower bound is given
in Ref.\cite{CF}
where the maximum likelihood equations were solved
iteratively and a formula for the covariance matrix of the ML estimators
was derived to one
higher order then given by the inverse of the Fisher information matrix.
This formula can be
treated as an approximation to the variances of the ML estimators by a
series in $1/d$ where $d$ is the signal-to-noise ratio.
The first order terms given by the inverse of Fisher matrix go as $1/d^2$
and the correction terms go like $1/d^4$. Consequently
one can expect that for signal-to-noise ratios of 10 or so the diagonal
elements of the
inverse of the Fisher matrix give variances of the ML estimators to
an accuracy of few \%.

We shall show that the set of parameters that we have chosen for our chirp
signal has particularly useful properties.
Note that the phase of the Fourier transform is linear in the phase,
the time parameter, and the
mass parameters $k_i$. We shall call these parameters {\em phase parameters}.
Moreover the Fourier transform is linear in the amplitude parameter
${\tilde A}$.
The maximum likelihood estimators are those values of the parameters
that maximize the likelihood ratio.
The expectation value of the log likelihood is given by
\be
E[\ln\Lambda] = (h|h_F) - \frac{1}{2}(h_F|h_F),
\ee
where $(h|h_F)$ is called the correlation function and is denoted by $H$.
Using the stationary phase approximation to the Fourier transform of
the signal $H$ is given by the integral
\begin{eqnarray}
& & H(\Delta t, \Delta\phi, \Delta k, \Delta k_e, \Delta k_1, \Delta k_{3/2},
\Delta k_2) = \\ \nonumber
& & 4 \tilde{A}\tilde{A}_F\int^{\infty}_{f_i}
\frac{df}{S_h(f) f^{7/3}}\cos[2\pi f\Delta t - \Delta\phi +
\frac{5}{48}(a(f;f_a)\Delta k + a_e(f;f_a)\Delta k_e \\ \nonumber
& & a_1(f;f_a)\Delta k_1 + a_{3/2}(f;f_a)\Delta k_{3/2} +
a_2(f;f_a)\Delta k_2)],
\label{cor}
\end{eqnarray}
where $\Delta t$ means the difference in time parameters of the signal
and the filter.
The expectation of the log likelihood ratio depends on the phase parameters
only through the correlation integral since
$(h_F|h_F) = H(0,0,0,0,0,0,0) = d^2$
where $d$ is the signal-to-noise ratio.
We see that the correlation function depends only on the differences
between the values of the phase parameters in the signal
and the filter and it has the maximum when the differences are zero.
Moreover the value of the correlation is the same if we move by the
same amount from the maximum in any direction for a given parameter
i.e. $H(-\Delta t,0,0,0,0,0,0) = H(\Delta t,0,0,0,0,0,0)$ and so on
for all phase parameters\footnote{We are indebted to Dr. J.A. Lobo
for this observation, see also \cite{He} p. 276.}.
This property means that the probability distribution of any estimator
of the phase parameter will be an even function of the difference between the
estimator and its true value. In other words the probability
distributions of the estimators of the phase parameters are symmetric
about their true values. Consequently we have
\be
m_l := E[(\hat{\theta_I} - \theta_I)^l] = 0 \hspace{5mm} \mbox{for l odd}
\label{mom}
\ee
and moreover for l even the moments $m_l$
are independent of the true values of the phase parameters.
Thus the ML estimators of the phase parameters are {\bf unbiased}
(this is immediate from Eq.\ref{mom} for $l = 1$) and the covariance matrix
of the estimators  of the phase parameters is independent of their values.
The probability distributions of the phase parameters will depend on
the signal-to-noise ratio. We know that for large signal-to-noise ratio
they will tend to Gaussian probability distributions.
The estimator of the amplitude parameter {\bf is biased} nevertheless by the
symmetry property of the probability distributions of the phase parameters
its bias is independent of the values of the phase parameters.
These properties of the parameters can be also seen explicitly
from the first two terms of the series
solution of the ML equations (Eq.\ref{ml}) given in ref.\cite{CF}.
The properties of our chosen set of parameters greatly simplify
calculation of the Cramer-Rao bounds.
In our case the Fisher information matrix $\Gamma$ is given by
\begin{equation}
\Gamma^{ij} = \frac{\partial H}{\partial\theta_{iS}
\partial\theta_{jF}}_{\theta_{kS}=\theta_{kF}},
\end{equation}
where $S$ refers to the parameters of the signal
and $F$ to the parameters of the filter.
The inverse of the ${\bf\Gamma}$ matrix is called the covariance matrix
and is denoted by ${\bf C}$.
It is easily seen that $\Gamma^{Ai}$ components are all equal to zero
when $i \neq A$. Thus the amplitude parameter decouples from the phase
parameters.
Because the phase parameters are unbiased the lower bounds of their
variances are given just by the appropriate diagonal elements of the
covariance matrix ${\bf C}$.
In the case of the amplitude parameter the Cramer-Rao bound is given by
$VarA \geq  b'(A)/\Gamma^{AA}$
where $b'(A)$ is the derivative of the bias of amplitude parameter w.r.t.
amplitude and $\Gamma^{AA} = d^2/\tilde{A}^2$.
Note that $\Gamma^{AA}$ is independent of $\tilde{A}$.
This is a consequence  of the linearity of the signal in the amplitude.

It is clear from the linearity of the function H in the differences
$\Delta\theta$
that the ${\bf \Gamma}$ matrix is independent
of the values of the phase parameters.
Thus the Cramer-Rao bound on these parameters is also
independent of the values of the parameters. From the argument above we
know that this holds not only for the bounds on the variances
but also for the variances themselves.

To obtain the maximum of the correlation
each phase parameter of the filter has to match
a corresponding parameter in the signal (see Eq.\ref{cor}). Thus
by linear filtering we shall get estimates of the time parameter $t_a$, phase,
and the mass parameters $k_i$.
In the filter one can always make an arbitrary choice
of the time parameter $t_a$.
For example instead of choosing $t_a$ as the time at which frequency
is $f_a$ one
can choose time $t'_a$ as the time at which the frequency is equal
to $f'_a$.
This new choice is equivalent to the following transformation
\begin{eqnarray}
t'_a = t_a + \bar{\delta}_0 k_0 + \bar{\delta}_e k_e + \bar{\delta}_1 k_1
+ \bar{\delta}_{3/2} k_{3/2} + \bar{\delta}_2 k_2, \\
\phi' = \phi + \delta_0 k_0 + \delta_e k_e + \delta_1 k_1
+ \delta_{3/2} k_{3/2} + \delta_2 k_2,
\label{tp1}
\end{eqnarray}
where
\begin{eqnarray}
\bar{\delta}_0 &=& \frac{5}{256}(\frac{1}{(\pi f_a)^{8/3}}
- \frac{1}{(\pi f'_a)^{8/3}}), \\
\delta_0 &=& \frac{1}{16}(\frac{1}{(\pi f_a)^{5/3}}
- \frac{1}{(\pi f'_a)^{5/3}}), \\
\bar{\delta}_e &=& \frac{785}{11008}(\frac{1}{(\pi f_a)^{43/9}}
- \frac{1}{(\pi f'_a)^{43/9}}), \\
\delta_e &=& \frac{785}{4352}(\frac{1}{(\pi f_a)^{34/9}}
- \frac{1}{(\pi f'_a)^{34/9}}), \\
\bar{\delta}_1 &=& \frac{5}{192}(\frac{1}{(\pi f_a)^2}
- \frac{1}{(\pi f'_a)^2}), \\
\delta_1 &=& \frac{5}{48}(\frac{1}{\pi f_a}
- \frac{1}{(\pi f'_a)}), \\
\bar{\delta}_{3/2} &=& \frac{1}{32}(\frac{1}{(\pi f_a)^{5/3}}
- \frac{1}{(\pi f'_a)^{5/3}}), \\
\delta_{3/2} &=& \frac{5}{32}(\frac{1}{(\pi f_a)^{2/3}} -
\frac{1}{(\pi f'_a)^{2/3}}), \\
\bar{\delta}_2 &=& \frac{5}{128}(\frac{1}{(\pi f_a)^{4/3}}
- \frac{1}{(\pi f'_a)^{4/3}}),\\
\delta_2 &=& \frac{5}{16}(\frac{1}{(\pi f_a)^{1/3}}
- \frac{1}{(\pi f'_a)^{1/3}}).
\label{tp}
\end{eqnarray}
The mass parameter frequency functions $a_i(f;f_a), (i=0,1,3/2,2)$
in Eq.\ref{Fou} are then transformed to $a_i(f;f'_a)$.
The mass parameters remain invariant under the above transformations.
By linear filtering with the template parametrized by the new time parameter
and the new phase given by the above transformation
we estimate the new time parameter $t'_a$ and the new phase $\phi'$
but the same mass parameters $k_i$.

There is also a particularly simple parametrization of the signal.
Let us rewrite the Fourier transform
of the gravitational wave signal from a binary in the following form
\bea
\hspace{-15mm}
\tilde{h} = \tilde{A} f^{-7/6} \exp i[2\pi f t_{c} - \phi_c - \pi/4  +
\\ \nonumber
\frac{3}{128}\frac{k}{(\pi f)^{5/3}} -
\frac{4239}{11696}\frac{k_e}{(\pi f)^{34/9}} +
\frac{5}{96}\frac{k_1}{\pi f} -
\frac{3}{32}\frac{k_{3/2}}{(\pi f)^{2/3}}
+ \frac{15}{64}\frac{k_2}{(\pi f)^{1/3}}],
\eea
(for $f > 0$ and by the complex conjugate
of the above expression for $f < 0$) where
$t_c$ and $\phi_c$ are coalescence time and phase respectively
and they are given by
\be
t_c = t_a + \frac{5}{256}\frac{k}{(\pi f_a)^{8/3}} -
\frac{785}{110008}\frac{k_e}{(\pi f)^{43/9}}
+ \frac{5}{192}\frac{k_1}{(\pi f_a)^2} -
\frac{1}{32}\frac{k_{3/2}}{(\pi f_a)^{5/3}}
+ \frac{5}{128}\frac{k_2}{(\pi f_a)^{4/3}}, \label{tc}
\ee
\be
\phi_c = \phi_a + \frac{1}{16}\frac{k}{(\pi f_a)^{5/3}} -
\frac{785}{4352}\frac{k_e}{(\pi f)^{34/9}} +
\frac{5}{48}\frac{k_1}{\pi f_a} -
\frac{5}{32}\frac{k_{3/2}}{(\pi f_a)^{2/3}} +
\frac{5}{16}\frac{k_2}{(\pi f_a)^{1/3}}. \label{pc}
\ee
Coalescence time  amd coalescence phase are obtained when the time parameter
$t_a'$ is such that the corresponding frequency $f_a'$
is infinite which occurs when the two point masses coalesce.
We can estimate the coalescence time and the coalescence phase of the
template if we filter for combinations of the time and phase parameters
with the mass parameter given precisely
by right hand sides of Eqs.(\ref{tc}) and (\ref{pc}).
There is also a transformation of the phase that we shall find useful
(see next Section).
\begin{equation}
\phi'' = \phi - 2 \pi f_m t_a .
\label{ph}
\end{equation}
where $f_m$ is some arbitrary constant frequency.
Using the new phase parameter in the filter given by above transformation
we shall estimate a new value of the phase shifted by
the amount $2 \pi f_m t_a$.
It is not difficult to show that all the above transformations do not change
the CR bound on the mass parameters however the transformation
Eq.(\ref{tp}) changes the bound
for time and phase parameters whereas
transformation Eq.(\ref{ph}) changes the bound on the phase.
We can use the freedom of these transformations in the filter
to obtain better accuracies of estimation of the time and the phase parameters.

\subsection{Numerical analysis of the rms errors of the estimators}

First of all we investigate the influence of the increasing number of post-
Newtonian parameters on the accuracy of their estimation.
To this end we have calculated the covariance matrices for the signal
containing only the quadrupole term, then covariance matrices
for 1st post-Newtonian, 3/2 post-Newtonian, and 2nd post-Newtonian signal,
and finally for the 2nd post-Newtonian signal with first order contribution
due to eccentricity.
The results are summarized in Table II where we have given rms errors
of the phase parameters.
We have given the rms errors for the time and phase of coalescence $t_c$
and $\phi_c$ respectively.
We have also determined the frequency $f_{min}$ for which the error
in the time parameter is minimum and we have given the minimum error
$\Delta t_{min}$ in the time parameter and the corresponding error
$\Delta\phi$
in phase.
We have considered a reference binary  of ${\cal M} = 1M_{\odot}$ located
at the distance of 100Mpc. We have taken the range of integration from
10Hz to infinity.
The signal-to-noise ratio for such a binary is around 25.

\bigskip

{\bf Table II.}
{\em The rms errors for the phase parameters at various post-Newtonian orders
for a reference binary of chirp mass of 1 solar mass at the distance
of 100Mpc. Expected advanced LIGO noise spectral density is assumed and
the integration range from 10Hz to infinity is taken giving signal-to-noise
ratio of around 25.}

\bigskip

{\footnotesize
\begin{tabular}{|c|c|c|c|c|c|c|c|c|} \hline
$\Delta t_m$[msec] & $\Delta\phi$ & $\Delta t_c$[msec] & $\Delta\phi_c$ &
$\Delta k [M_{\odot}^{-5/3}]$ & $\Delta k_1 [M_{\odot}^{-1}]$
& $\Delta k_{3/2} [M_{\odot}^{-2/3}]$ & $\Delta k_2 [M_{\odot}^{-1/3}]$ &
$\Delta k_e [M_{\odot}^{-5/3}$100Hz$^{19/9}]$ \\ \hline
0.14 & 0.073 & 0.17 & 0.10 & 8.3 $\times 10^{-6}$ & - & - & - & - \\
0.15 & 0.087 & 0.27 & 0.33 & 4.0 $\times 10^{-5}$ & 5.8
$\times 10^{-3}$ & - & - & - \\
0.18 & 0.14 & 0.54 & 1.9 & 1.7 $\times 10^{-4}$ &  0.70
$\times 10^{-1}$ & 0.52 & - & - \\
0.24 & 0.14 & 1.6 & 24 & 6.6 $\times 10^{-4}$ &  0.50 & 7.2 & 28 & - \\
0.25 & 0.17 & 2.3 & 45 & 2.3 $\times 10^{-3}$ &  1.3 & 17 & 59 & 1.2
$\times 10^{-6}$ \\ \hline
\end{tabular}}

\bigskip

The above bounds scale exactly as the inverse of the signal-to-noise ratio d
and they do not depend on the numerical values of the phase parameters.
 From the above table we see that increasing the number of post-Newtonian
corrections and parameters
we filter for decreases the accuracy of estimation of the parameters
independently of the size of the post-Newtonian correction. Thus searching
for a negligible correction due to eccentricity increases the rms error in
other parameters by over 100\%.

For completeness in Appendix B we give the numerical values
of covariance matrices
for the phase parameters at various post-Newtonian orders and the
corresponding values of the frequency $f_{min}$.

As we have indicated above the estimator of the amplitude parameter
is biased however
if one takes the expansion of the variance of the estimator in the inverse
powers of the signal-to-noise ratio (see \cite{CF} for a general formula)
then the leading term for the variance of the amplitude is just
$1/\Gamma^{AA}$ where $\Gamma^{AA}$ is independent of $\tilde{A}$.
The higher order corrections to the CR bounds of the amplitude
go like $1/d^4$ and they do depend on the value of the amplitude.
As an amplitude parameter we find convenient to choose $A_{\oplus}$ given by
\begin{equation}
A_{\oplus} = \frac{{\cal M}_{\odot}^{5/6}}{r_{100Mpc}}
\end{equation}
where $\cal M_{\odot}$ is the chirp mass in the units of solar masses
and $r_{100Mpc}$ is the distance in the units of 100Mpc.
For our reference binary the amplitude $A_{\oplus} = 1$ and
thus the approximate rms error in its  ML estimator is
$A_{\oplus}/d \simeq 1/25 = 0.04$ and as explained above this last number
is independent of the true value of the amplitude.

It is important to assess the accuracy of estimation of the physical parameters
of the binary, i.e. the two masses of its members and the spin parameters
$s_o$ and $s_s$.
This means that we have to make a transformation to a different parameter set.
A nice property of the ML estimators is the following.
Let $\hat{\theta}_i$ be the maximum likelihood estimators of the set
of parameters ${\theta_i}$. Let $f(\theta_i)$ be a function of the parameters
then $f(\hat{\theta}_i)$ is the maximum likelihood estimator of the function
$f$ (see Ref\cite{D}).
However it is not true in general that if estimators of the old parameters are
unbiased then the new parameter is unbiased as well.
Consequently by transforming the bounds of the old
parameters one will not get the Cramer-Rao bound on the new set of parameters.
However we know that Cramer-Rao bounds are approximately equal to the true
variances in the limit of high signal-to-noise ratio $d$, correction terms
being of the order of $1/d^2$. Hence by transforming the
C-R bounds one gets the rms errors of the estimators
accurate to the order $1/d$.
Another important point is that the transformation to the new parameter set
may be singular.
Then the determinant of the ${\bf\Gamma}'$ matrix for the new set of parameters
is zero  and thus ${\bf\Gamma}'$ is not positive definite, consequently
the Cramer-Rao inequality does not hold. A way to get errors of estimators
of the new parameters in such a case could be to attempt to calculate
the bias and the variance directly from
some approximate probability distributions for the estimators
(see ref.\cite{CF} for such treatment to determine the accuracy
of the distance to the binary). It may happen however that the
probability density function is such that the expectation value and the
variance
do not exist (an example is Cauchy probability distribution) and
then one may have to use another measure of bias and error, e.g. median
and interquartile distance.
The other method proposed in \cite{CF} is to use confidence intervals.
We shall return to this problem in the future work \cite{JKKT,KL}.

The transformation from the 4 mass parameters $k_I$ to new parameters -
total mass (m) , reduced mass ($\mu$) and the spin parameters $s_o$ and $s_s$
is regular. Thus we can obtain approximate values of the errors
of the estimators of the reduced mass, the total mass and the spin parameters.
However the transformation from m and $\mu$ to individual masses
$m_1$ and $m_2$  is singular
(determinant of the Jacobian of the transformation is zero
when masses are equal, see Ref.\cite{CF}).
Consequently the errors in the determination of the masses
cannot be obtained from the C-R bounds calculated above.

In Table III we show the degradation of the accuracy of estimation
of the chirp mass, the reduced mass, and the total mass with
the increasing number
of parameters in the signal for the NS-NS binary at a distance of 200Mpc.

\bigskip

{\bf Table III.}
{\em Degradation of the accuracy of estimation of the chirp mass,
the reduced mass, and the total mass for the fiducial
neutron star - neutron star binary with increasing number of parameters
in the signal. 2 pN means that the phase of signal is taken to 2nd post-
Newtonian order with spin parameters included and we maximize the correlation
of the signal with a template matched to the signal
for all the phase parameters.}

\bigskip

\begin{center}
\begin{tabular}{|c|c|c|c|} \hline
pN order & $\Delta {\cal M}/{\cal M}$ & $\Delta \mu/\mu$ & $\Delta m/m$
\\ \hline
1 pN & 0.0054\% & 0.55\% & 0.81\% \\
3/2 pN & 0.023\% & 6.4\% & 9.6\% \\
2 pN & 0.080\% & 42\% & 63\% \\ \hline
\end{tabular}
\end{center}

\bigskip

For the calculation of the numbers in the table above and all other tables
in the remaining part of this Section we have taken the range of integration
in the Fisher matrix integrals to be from 10Hz
to the frequency $f = (6^{3/2}\pi m)^{-1}$ corresponding to the last
stable orbit of the test particle in Schwartzschild space-time. This may very
roughly correspond to the last stable orbit in a binary \cite{WS,KWW2}.

In Table IV we give the signal-to-noise ratios and
the Cramer-Rao bounds for the mass and the spin parameters in percents of
their true values for the 2nd post-Newtonian signal
for our three representative binary systems
at the distance of 200Mpc. We have also given the improvement factors
in $\sqrt{n}$ in the S/N due to filtering.

{\bf Table IV.}
{\em Accuracy of estimation of the parameters of the 2nd post-Newtonian
signal for the three fiducial binaries.}

\bigskip

\begin{tabular}{|c|c|c|c|c|c|c|c|c|} \hline
Binary & S/N & $\sqrt{n}$ & $\Delta {\cal M}/{\cal M}$
& $\Delta \mu/\mu$ & $\Delta m/m$ & $\Delta s_o/s_o$ & $\Delta s_s/s_s$
\\ \hline
NS-NS  & 15 & 32 & 0.080\% & 42\% & 63\%
& $60 \times 10^2$\% & $12 \times 10^6$\% \\
NS-BH  & 32 & 15 & 0.26\% & 40\% & 59\% & 11\% & $19 \times 10^4$\% \\
BH-BH  & 77 & 6 & 0.92\% & 150\% & 230\% & 240\% & 890\% \\ \hline
\end{tabular}

\bigskip

We see that only the rms error in the chirp mass is small and also the
accuracy of the determination of the spin-orbit parameter
for NS-BH binary is satisfactory. The errors in reduced and total masses
are large.

One can derive simple general formulae for the accuracy of determination
of the chirp mass, the reduced mass, and the total mass in terms of rms errors
of the mass parameters $k_i$. From the definition of the chirp mass one
immediately obtains the following
formula for the relative rms error in terms of the rms error in the mass
parameter k,
\be
\Delta {\cal M}/{\cal M} =
\frac{3}{5} r_{100Mpc} \Delta k {\cal M}_{\odot}^{5/3}.
\ee
For the errors in the reduced and the total mass we obtain
the following general formulae using the standard law of propagation of errors
\bea
\Delta \mu = \frac{| - \frac{\partial k_1}{m}\sqrt{\Delta k} +
\frac{\partial k}{m}\sqrt{\Delta k_1}|}{det}, \\
\Delta m = \frac{|\frac{\partial k_1}{\mu}\sqrt{\Delta k} -
\frac{\partial k}{\mu}\sqrt{\Delta k_1}|}{det},
\eea
where
\be
det = \frac{\partial k}{m}\frac{\partial k_1}{\partial\mu} -
\frac{\partial k}{\partial\mu}\frac{\partial k_1}{\partial m}
\ee
and $\Delta k$, $\Delta k_1$ are rms error in mass parameters $k$ and $k_1$
respectively.
The formula above is the same when the 1st post-Newtonian,
the 3/2 post-Newtonian, and the 2nd post-Newtonian corrections are included.
We observe that errors in $\mu$ and $m$ depend only on the masses and the rms
errors in the parameters $k$ and $k_1$. The other mass parameters influence
the errors in $\mu$ and $m$ only through their correlations with the mass
parameters $k$ and $k_1$ and only through the functional
form of the corrections as the rms error in the mass parameters are
independent of their values.
The errors in $\mu$ and $m$ are independent
of the numerical values of the parameters $k_{3/2}$ and $k_2$.
Since in general the rms error $\Delta k$ is considerably smaller than
$\Delta k_1$ we get the following simplified expressions for the relative
errors in the reduced and the total mass.
\bea
\Delta \mu/\mu = \frac{1}{a} r_{100Mpc}\mu_{\odot}\Delta k_1, \label{b} \\
\Delta m/m = \frac{3}{2 a} r_{100Mpc}\mu_{\odot}\Delta k_1, \label{c}
\end{eqnarray}
where a = 743/336 - 33/8 $\mu/m$.
We see that the error in the determination of the reduced mass
and the total mass
is determined by error in the first post-Newtonian mass parameter $k_1$.
Since the ratio $\mu/m$ is $\leq 1/4$
to a fairly good approximation we can take the value of $a$ roughly equal to 1.

If the spin effects could entirely be neglected and we would only have
the reduced mass and the total mass as unknown in the mass parameters $k_i$
then we could achieve the accuracies in the parameters of the signal
summarized in Table V.  We considered three fiducial binary systems
and 2nd post-Newtonian signal but with spin-orbit
and spin-spin parameters removed.
Thus the number of parameters estimated is 2 less than for the signal
considered in Table IV.

{\bf Table V}
{\em Accuracy of estimation of the parameters for 3 fiducial
binary systems and the 2nd post-Newtonian signal but with spin parameters
removed.}

\bigskip

\begin{center}
\begin{tabular}{|c|c|c|c|c|c|} \hline
Binary & S/N & $\Delta t_c$ms & $\Delta\phi_c$
& $\Delta \mu/\mu$ & $\Delta m/m$ \\ \hline
NS-NS  & 15 & 0.47 & 0.82 & 0.29\% & 0.43\% \\
NS-BH  & 32 & 0.32 & 0.47 & 0.19\% & 0.28\% \\
BH-BH  & 77 & 0.18 & 0.24 & 0.27\% & 0.37\% \\ \hline
\end{tabular}
\end{center}
\bigskip

We see that if spin parameters could be neglected we would have an
excellent accuracy of estimation of the reduced and the total mass
of the binary.

\subsection{The effects of other theories of gravity}

We shall consider two alternative theories.
One is Jordan-Fiertz-Brans-Dicke (JFBD) theory
(see \cite{W} for a detailed discussion) and the other is a multi-scalar field
theory recently proposed in \cite{DEF}.

In the JFBD theory in addition to the tensor gravitational field there is
also a scalar
field. The theory can be characterized by a coupling constant that we
denote by $\omega$.
General relativity is obtained when $\omega$ goes to infinity.
The JFBD theory has two effects on gravitational emission.
It admits dipole gravitational radiation and secondly there is a modification
of the quadrupole emission due to the interaction of the scalar field with
gravitating
bodies. In the case of binary system the effects of the JFBD theory
has been studied
in great detail \cite{WZ} and a general formula for the change of orbital
period was derived (\cite{W} eq.(14.22)). From that formula we
get the following expression for the characteristic time $\tau$
of the evolution of the binary due to radiation reaction in the case of
circularized orbits
and assuming that the contribution due to the dipole term is small
\begin{eqnarray}
\tau =
\frac{5}{96}\frac{1}{\mu m^{2/3}}\frac{{\cal G}^{4/3}}{\kappa}\frac{1}
{(\pi f)^{8/3}} \times \\ \nonumber
(1 - \frac{5}{192} k_B \frac{{\cal G}^{4/3}}{\kappa}\frac{\Sigma^2}
{(\pi m f)^{2/3}}),
\end{eqnarray}
where
\begin{eqnarray}
k_B &=& \frac{1}{2 + \omega}, \\
{\cal G} &=& 1 - \frac{k_B}{2} ({\cal C}_1 + {\cal C}_2 - {\cal C}_1
{\cal C}_2), \\
\kappa &=& {\cal G}^2 (1 - \frac{k_B}{2} + \frac{k_B}{12}\gamma^2), \\
\gamma &=& 1 - \frac{m_1 {\cal C}_2 + m_2 {\cal C}_1}{m_1 + m_2}, \\
\Sigma  &=& {\cal C}_1 - {\cal C}_2.
\end{eqnarray}
${\cal C}_1$ and ${\cal C}_2$  are ``sensitivities" of the two bodies
to changes of the scalar field.
For a black hole the sensitivity ${\cal C}$  is always equal to 1.
For a neutron star ${\cal C}$
depends on the equation of state. For neutron stars the sensitivity has
been studied in \cite{DEF}
for a number of equations of state and it was found for a wide range of
such equations
that it is proportional to the mass of the neutron star with proportionality
constant varying
from .17 to .31. Here we shall assume that
${\cal C}_i = 0.21 m_{i\odot}$ for a neutron star
of $m_{i\odot}$ solar mases. From the above formulae one sees that the
dipole radiation will vanish if the binary system
consists of two black holes or the neutron stars in the binary are the same.

The Fourier transform of the signal in the stationary phase approximation
including contributions due to JFKB theory is given by (we neglect any
contributions due to eccentricity)
\begin{eqnarray}
& &\tilde{h} = \tilde{A} f^{-7/6} \exp i[2\pi f t_{a} - \phi - \pi/4
\\ \nonumber
& &+ \frac{5}{48}(a(f;f_a) k' + a_1(f;f_a) k_1 + a_{3/2}(f;f_a) k_{3/2}
+ a_2(f;f_a) a_d(f;f_a)) k_d)]
\end{eqnarray}
for $f > 0$ and by the complex conjugate of the above expression for $f < 0$
where the function $a_d(f;f_a)$ due to dipole radiation has the form
\begin{eqnarray}
a_d(f;f_a) = - \frac{5}{192} (\frac{9}{70}\frac{1}{(\pi f)^{7/3}} +
\frac{3}{10}\frac{f}{(\pi f_a)^{10/3}} - \frac{3}{7}\frac{1}{(\pi f_a)^{7/3}}),
\end{eqnarray}
and where
\begin{eqnarray}
k' &=& \frac{1}{\mu m^{2/3}}\frac{{\cal G}^{4/3}}{\kappa}, \\
k_d &=& \frac{1}{\mu m^{4/3}} k_B \frac{{\cal G}^{8/3}}{{\kappa}^2}\Sigma^2
\end{eqnarray}
Curent observational tests constrain $\omega$ to be greater
than 600 and from timing of binary pulsar a lower limit on $\omega$
of 200 can be set. Thus it is sufficient to keep only
the first terms in $1/\omega$. Then the two parameters above are approximately
given by
\begin{eqnarray}
k' &=& \frac{1}{\mu m^{2/3}}\,(1 - dk_d), \\
k_d &=& \frac{1}{\mu m^{4/3}}\,k_B {\Sigma}^2
\end{eqnarray}
where
\begin{equation}
dk_d = k_B [\frac{1}{3} ({\cal C}_1 + {\cal C}_2 -
{\cal C}_1 {\cal C}_2) + \frac{1}{2} - \frac{\gamma^2}{12}].
\end{equation}
We thus see the the JFBD theory introduces a new parameter
$k_d$ due to the dipole radiation and
modifies the standard chirp mass parameter k by fraction $dk_d$.

We have investigated the potential accuracy of estimation of the
parameter $k_d$ assuming that the spin effects are negligible.
We have taken neutron star/black hole binary with parameters given in
Table I at the distance of 200Mpc.
The result is summarized in Table VI.

{\bf Table VI.}
{\em The rms error for signal parameters in JFKB theory assuming spins
are negligible for the binary of 1.4 solar mass neutron star and 10 solar mass
black hole.}

\bigskip

\begin{center}
\begin{tabular}{|c|c|c|c|c|c|} \hline
S/N & $\Delta t_c$[ms] & $\Delta\phi_c$
& $\Delta \mu/\mu$ & $\Delta m/m$ & $\Delta k_d [M_{\odot}^{-5/3}]$ \\ \hline
32 & 0.47 & 0.97 & 0.57\% & 0.73\% & $2.3\times10^{-5}$ \\ \hline
\end{tabular}
\end{center}

\bigskip

The potential accuracy of determination of the dipole radiation
parameter $k_d$ is high. Current observational constraints indicate
however that this parameter is small. We have the following numerical
values.
\bea
k_d = 3.2\times 10^{-5} (\frac{500}{\omega}) (\frac{\Sigma^2}{0.5})
(\frac{32}{\mu_{\odot}m^{4/3}_{\odot}}) \\
\frac{\Delta k_d}{k_d} = 0.7 (\frac{\omega}{500}) (\frac{0.5}{\Sigma^2})
(\frac{\mu_{\odot} m^{4/3}_{\odot}}{32})
\eea
We conclude that the gravitational-wave measurement by planned long arm
laser interferometers have the potential of testing JFBD to the
accuracy comparable to tests in solar system and measurements from
the binary pulsars \cite{Will}.

 From the general class of tensor-multi-scalar theories studied
recently \cite{DEF} we shall
consider a two-parameter subclass of tensor-bi-scalar theories denoted by
T($\beta'$,$\beta''$). Theories in this subclass have two scalar fields
and they tend smoothly
to general theory of relativity when both parameters
$\beta'$ and $\beta''$ tend to zero.
The subclass is defined in such a way that the dipole radiation
vanishes. From the general formulae \cite{DEF} one can calculate
the characteristic time $\tau$. For circularized orbits the only
modification is an
effective change of the chirp mass parameter k given by the following formula
\begin{eqnarray}
k'' & = & k - d_{DF}, \\
d_{DF} & = & \frac{5}{144}\kappa_o(m_1,{\cal C}_1,m_2,{\cal C}_2) +
\frac{1}{6}(\kappa_q(m_1,{\cal C}_1,m_2,{\cal C}_2) + \\ \nonumber
& & \kappa_{d1}(m_1,{\cal C}_1,m_2,{\cal C}_2)) +
\frac{5}{48}\kappa_{d2}(m_1,{\cal C}_1,m_2,{\cal C}_2)
\end{eqnarray}
where coefficients $\kappa_o, \kappa_q, \kappa_{d1}, \kappa_{d2}$
are due to contributions from
quadrupole helicity zero, corrections to quadrupole helicity two,
and dipole radiation respectively.
They are complicated functions of the masses and sensitivities.
We give the detailed formulae in Appendix C.
In all the tensor-multi-scalar field theories
whenever one of the component is a black
hole corrections to the radiation reaction vanish. We have also found that for
a simple model where sensitivities are proportional
to masses of neutron stars and the proportionality constant
is the same the correction $d_{DF}$ does not depend on the parameter
$\beta''$. For a system of two identical neutron
stars the correction $d_{DF}$ takes a simple form
\begin{equation}
d_{DF} = 0.21 \beta {\cal C}^2,
\label{sh}
\end{equation}
where ${\cal C}$ is the sensitivity of  the neutron star to changes
of the scalar field introduced above.
Current observations constrain parameter $\beta$ to be less than 1.
For circularized orbits (the case considered above) the bi-scalar theory
does not introduce a new mass parameter in the phase of the signal but only
a shift in the ``Newtonian" mass parameter $k$. We shall consider
the possibility of estimating this shift in the next section.

\section{Search templates}
\subsection{The Newtonian filter}

We have seen in the previous Section that
the accuracy of estimation of the parameters is significantly degraded
with increasing number of corrections even though a correction may be small.
If we include the 2nd post-Newtonian correction and filter for all
unknown parameters then the accuracy of determination of the masses of the
binary becomes undesirably low.
Moreover we cannot entirely exclude unpredicted small effects
in the gravitational-wave emission
(e.g. corrections to general theory of gravity)
that we at present cannot model. Thus there is a need for simple filters
or {\em search templates} that will enable us to scan the data effectively
and isolate stretches of data where the signal is most likely
to be \cite{Cal}. The simplest such filter is just a Newtonian waveform
$h_N$ which Fourier transform in stationary phase approximation is given by
\be
\tilde{h}_N = \frac{1}{30^{1/2}}\frac{1}{\pi^{2/3}}\frac{\mu^{1/2} m^{1/3}}{R}
f^{-7/6} \exp i[2\pi f t_c - \phi_c - \pi/4 + k\frac{3}{128}(\pi f)^{-5/3}].
\ee
We shall call the {\em Newtonian filter} the filter which
Fourier transform is given by the above formula and we shall denote it
by $Nf$.
This filter has been investigated by the present authors \cite{KKT,Kc,KKS}
and also by other researchers \cite{FlN,FN,BD,S,Apo}.
A different search template based on the post-Newtonian signal
has recently been introduced in Ref.\cite{Apo}. We discuss this
alternative search template in the next subsection.

In this section we examine the performance of the Newtonian filter.
We demonstrate that
such a template will perform well in detecting the signal
from a binary and it also gives a reasonable idea of the nature of the binary.
We shall investigate the performance of the Newtonian filter both
analytically and numerically.

Let us consider the correlation of the post-Newtonian signal
with the Newtonian filter. Such an integral has the same form as
the correlation integral given by Eq.\ref{cor} in Section 2.1 except
that all post-Newtonian mass parameters will be unmatched by the
parameters of the filter.
The correlation will be high if we can reduce the oscillations due to
the cosine function as much as possible.
Since the integrand of the correlation integral is fairly sharply peaked
($HWHM \simeq 26$Hz) around its maximum at the frequency $f'_o \simeq 47$Hz
we can achieve this by making the phase as small as
possible around the peak frequency $f'_o$.
The argument $\Phi$ of the cosine in the integrand of the correlation
of the post-Newtonian signal with the Newtonian filter including the effects
due to eccentricity and dipole radiation takes the form
\bea
& &\Phi(f) = 2\pi f\Delta t + \Delta\phi + \\ \nonumber
& &\frac{5}{48}[a(f;f_a)\Delta k + a_e(f;f_a) k_e  +
a_1(f;f_a) k_1 + a_{3/2}(f;f_a) k_{3/2} + a_2(f;f_a) k_2 +
a_d(f;f_a) k_d].
\eea
First we note that for all the mass parameter
frequency functions $a_i(f;f_a)$
the functions and their first derivatives vanish at the frequency $f_a$.
We shall therefore choose $f_a = f'_o$. Let us also transform the phase
parameter according to transformation given by Eq.\ref{ph} with $f_m = f'_o$.
In the new parametrization the phase $\Phi$ takes the form
\bea
&&\Phi(f) = 2\pi (f - f'_o) \Delta t' + \Delta\phi'' + \\ \nonumber
&&\frac{5}{48}[a(f;f'_o)\Delta k - a_e(f;f'_o) k_e + a_1(f;f'_o) k_1
- a_{3/2}(f;f'_o) k_{3/2} + a_2(f;f'_o) k_2 - a_d(f;f'_o) k_d].
\label{Ph}
\eea
Let us examine the functional behaviour of $\Phi(f)$
around the  frequency $f'_o$. We find
\begin{eqnarray}
& &\Phi(f) \simeq 2\pi (f - f'_o) \Delta t' + \Delta\phi'' + \\ \nonumber
& &\frac{5}{96}(f/f'_o - 1)^2[\frac{\Delta k}{(\pi f'_o)^{5/3}} -
\frac{157}{24}\frac{k_e}{(\pi f'_o)^{19/9}} +
\frac{k_1}{\pi f'_o} - \frac{k_{3/2}}{(\pi f'_o)^{2/3}} +
\frac{k_2}{(\pi f'_o)^{1/3}} -
\frac{5}{192}\frac{k_d}{(\pi f'_o)^{7/3}}] \\ \nonumber
& & + \ O[(f/f'_o - 1)^3].
\end{eqnarray}
We see that in the above approximation we can make the phase $\Phi$
vanish to the order $(f/f'_o - 1)^3$ when the following conditions hold
\begin{eqnarray}
& & \Delta t_{max} = t'_{Fmax} - t' = 0,\\
& & \Delta\phi'' = \phi''_{Fmax} - \phi'' = 0, \\
& & \Delta k_{max} = k_{Fmax} - k = \\ \label{kn} \nonumber
& & - \frac{157}{24}\frac{k_e}{(\pi f'_o)^{19/9}} + k_1(\pi f'_o)^{2/3}
- k_{3/2}(\pi f'_o) + k_2(\pi f'_o)^{4/3}
- \frac{5}{192}\frac{k_d}{(\pi f'_o)^{2/3}}
\end{eqnarray}
where subscript $Fmax$ means the value of the parameter of the Newtonian
filter that maximizes the correlation.
Hence we can expect to match the Newtonian template
to the post-Newtonian signal
with the Newtonian mass parameter $k$ shifted from the true value
by a certain well-defined amount.
The shift depends both on the parameters of the two-body system and the
noise in the detector through the frequency $f'_o$.
However the value of the shift in the $k$ parameter is  independent
of the choice of the time parameter and phase in the Newtonian filter.

In the following table we have given the numerical values of the shift in the
parameter $k$ calculated from Eq.\ref{kn}
for the 3 binary systems considered in the previous section. We have
given three values of the shifts including one ($\delta k_1$),
two ($\delta k_{3/2}$), and finally three ($\delta k_2$)
post-Newtonian corrections.

\bigskip

{\bf Table VII.}
{\em Numerical values of the shifts in the mass parameter of the
Newtonian filter
calculated from the analytic formula (Eq.\ref{kn}).}

\bigskip

\begin{center}
\begin{tabular}{|c|c|c|c|} \hline
Binary & $\delta k_1$ & $\delta k_{3/2}$ & $\delta k_2$ \\ \hline
NS-NS & 0.03328 & 0.01512 & 0.01597 \\
NS-BH & 0.01641 & 0.005052 & 0.006023 \\
BH-BH & 0.004660 & 0.001276 & 0.001775 \\ \hline
\end{tabular}
\end{center}

\bigskip

We have also investigated the problem numerically and we have
found the maxima to be located at the values of the shifts in the phase,
the time, and the mass parameter $k$ given in Tables VIIIA (1st post-Newtonian
shift), VIIIB (3/2 post-Newtonian shift),
VIIIC (2nd post-Newtonian shift) below.
We have also given the factor $l$ which is defined as
\be
l = \sqrt{\frac{(h|h_N)}{(h|h)}}
\ee
In a previous work by these authors (\cite{Kc,KKS}) we have claimed the factor
$l$ to be the drop in the signal-to-noise ratio as a result of using
non-optimal
(Newtonian) filter. However the signal-to-noise ratio falls as {\em square}
of the factor $l$\footnote{We are grateful to T. Apostolatos for pointing this
to us.} (see Appendix D).
We also give the range of integration over which we calculated the correlation.
We have found that the we gain very little by extending the
integration beyond that range.
For the case of a neutron star binary increasing the range
of integration up to 800Hz increases the signal-to-noise ratio by less than
1\%.
The reason for this is the effective narrowing
of the band of the detector by the chirp signal
discussed in the previous section.

\newpage

{\bf Table VIII.}
{\em Numerical values of the factor l and shifts in the parameters of the
Newtonian filter
with respect to the true values for various post-Newtonian orders
calculated numerically by maximizing the correlation function.}

\bigskip

A

\begin{tabular}{|c|c|c|c|c|c|} \hline
Binary & $l_1$ &  $\delta k_1$ & $\delta t'$ & $\delta\phi''$
&  Range \\ \hline
NS-NS & 0.68 &  0.03721 & $3.0 \times 10^{-3}$ & 0.61 & 30Hz - 200Hz \\
NS-BH & 0.76 &  0.01867  & $1.7 \times 10^{-3}$ & -0.53 & 30Hz - 100Hz \\
BH-BH & 0.85 &  0.004931 & $-4.1 \times 10^{-3}$ & -0.20 & 30Hz - 100Hz
\\ \hline
\end{tabular}

\bigskip

B

\begin{tabular}{|c|c|c|c|c|c|} \hline
Binary & $l_{3/2}$  & $\delta k_{3/2}$ & $\delta t'$ & $\delta\phi''$
& Range \\ \hline
NS-NS & 0.90 & 0.01564 & $-3.5 \times 10^{-3}$ & -0.40 & 30Hz - 200Hz \\
NS-BH & 0.87 & 0.004905 & $0.61 \times 10^{-3}$ & 0.068 & 30Hz - 100Hz \\
BH-BH & 0.87 & -0.001219 & $0.39 \times 10^{-3}$ & 0.030 & 30Hz - 100Hz  \\
\hline
\end{tabular}

\bigskip

C

\begin{tabular}{|c|c|c|c|c|c|} \hline
Binary & $l_2$  & $\delta k_{2}$ & $\delta t'$ & $\delta\phi''$ & Range \\
\hline
NS-NS & 0.85 & 0.01658 & $-5.5 \times 10^{-3}$ & -0.44 & 30Hz - 200Hz \\
NS-BH & 0.87 & 0.006014 & -$1.1 \times 10^{-3}$ & -0.024 & 30Hz - 100Hz \\
BH-BH & 0.87 & 0.001789 & -$0.51 \times 10^{-3}$ & -0.018 & 30Hz - 100Hz  \\
\hline
\end{tabular}

\bigskip

We see that the agreement between the predicted values of the shifts in
the  parameters and the numerical values given above is very good. In
particular
the difference between the predicted values  and the values of the shifts
for the k parameter obtained numerically differ by less then 5\%.

The results of the detailed analysis carried out in \cite{Apo} show that
when the amplitude and phase modulations due to the time dependence
of the spin parameters are taken into account then
in the worst case $l = 0.63$ for the correlation of the Newtonian filter
with the 3/2pN signal.

We have also performed the correlation using the signal in the time domain
and evaluating the correlation using the fast Fourier transform.
We kept the amplitude Newtonian. As we have remarked earlier the restricted
post-Newtonian approximation are not equivalent in the frequency
and the time domain. So the results are not the same.

\newpage

{\bf Table IX.}
{\em Numerical values of the l factor and the shifts obtained from the
correlation of the Newtonian template with the signal in the time domain
at various post-Newtonian orders.}

\begin{center}
\begin{tabular}{|c|c|c|c|c|c|c|c|} \hline
Binary & $l_1$ & $\delta k_1$ & $l_{3/2}$ & $\delta k_{3/2}$ &
$l_2$ & $\delta k_2$ & Range \\ \hline
NS-NS & 0.67 & 0.04097 & 0.97 & 0.01576 & 0.88 & 0.01899 & 30Hz - 200Hz \\
NS-BH & 0.87 & 0.01916 & 1.00 & 0.004889 & 0.93 & 0.01097 & 30Hz - 100Hz \\
BH-BH & 0.97 & 0.005130 & 1.00 & 0.001896 & 0.94 & 0.005874 & 30Hz - 100Hz
\\ \hline
\end{tabular}
\end{center}

\bigskip

We therefore conclude that the Newtonian filter will perform reasonably well
in {\em detecting} the post-Newtonian signal.

Using the Newtonian filter we would not like to loose any signals.
We can achieve this by suitably lowering the detection
threshold when filtering the data with the Newtonian filter.
By this procedure we would isolate stretches of data
where correlation has crossed the lowered threshold.
The reduced data would contain all the signals that would be detected
with the optimal filter but would also contain false alarms
which number would be increased comparing to number of false alarms with
the optimal filter. This is the effect of lowering the threshold.
The next step would be to analyse
the reduced set of data with more accurate templates and the initial
threshold to make the final detection.

In Table X we have given examples of the performance of the above procedure.
We assume the signal-to-noise ratio threshold $d_T = 5$ and
we assume we have 1 signal for the optimal signal-to-noise ratio d.
N is the expected number of detected signals with the optimal filter,
$N_F$ is the number of false alarms,
$N_N$ is the number of detected signals with the Newtonian filter,
$T_N$ is the lowered threshold, $N_L$ is the number of
signals with the lowered
threshold  and $N_{FL}$ is the number of false alarms with the
lowered threshold
(see Appendix D for definition of these quatities).

\newpage

{\bf Table X.}
{\em Comparision of number of true events and false alarms obtained with
the optimal filter and the Newtonian filter.}

\bigskip

\begin{tabular}{|c|c|c|c|c|c|c|c|} \hline
d & FF & N & $N_F$ & $N_N$ & $T_N$ & $N_L$ & $N_{FL}$ \\ \hline
15 & .81 & 27 & 0.055 & 20 & 4.5 & 28 & 0.16 \\
15 & .36 & 27 & 0.055 & 5.6 & 3.225 & 28 & 2.1 \\
30 & .81 & 225 & 1.1 & 165 & 4.5 & 230 & 2.2 \\
30 & .25 & 225 & 1.1 & 31 & 2.875 & 229 & 32 \\ \hline
\end{tabular}

\bigskip

The theory of filtering with a suboptimal filter is outlined in Appendix D
and the terms used in this Section are precisely defined.

We have also calculated the covariance matrix for the parameters
estimated with the
Newtonian filter. Calculating the second derivatives of the
correlation function at the maximum given by the numerical values
of the parameters in Table VIII one gets the {\bf $\Gamma$} matrix.
The inverse gives the covariance matrix.
The square roots of a diagonal components of the covariance matrix
give lower bounds on the accuracy of determination
of parameters with the Newtonian filter and they are approximate
rms error for high signal-to-noise ratio as explained in Section 2.
The results are summarized in Table XI for our
three binary systems located at the distance of 200Mpc.
The numbers are given for signals with the currently known post-Newtonian
corrections but without the eccentricity and the dipole terms.

\bigskip

{\bf Table XI.}
{\em Accuracy of determination of parameters of the Newtonian filter
for the three fiducial binaries located at the distance of 200Mpc.}

\bigskip

\begin{center}
\begin{tabular}{|c|c|c|c|c|c|c|c|c|} \hline
Binary & $\Delta t_{aN}$[ms] & $\Delta k_N [M_{\odot}^{-5/3}]$ \\ \hline
NS-NS  & 2.9 & $0.37 \times 10^{-3}$ \\
NS-BH  & 0.53 & $0.051 \times 10^{-3}$ \\
BH-BH  & 0.22 & $0.021 \times 10^{-3}$ \\ \hline
\end{tabular}
\end{center}

\bigskip

One can easily calculate from Table II that the accuracy of determination of
the mass parameter $k$ with the Newtonian filter lies between the accuracy
of determination of $k$ for 1 and 3/2 post-Newtonian signal.

In Appendix E we have derived a useful formula for the correlation
function based on the approximation to the phase $\Phi$ considered above.

We shall next show that the Newtonian filter can also give
a useful estimator characterizing the binary system.
 From the analytic investigation of the Newtonian filter given above
it is clear that we can obtain an estimator of an {\em effective mass
parameter} $k_E$ of the binary system given approximately by (cf.Eq.\ref{kn})
\be
k_E = k - \frac{157}{24}\frac{k_e}{(\pi f'_o)^{19/9}} + k_1(\pi f'_o)^{2/3}
- k_{3/2}(\pi f'_o) + k_2(\pi f'_o)^{4/3}
- \frac{5}{192}\frac{k_d}{(\pi f'_o)^{2/3}}
\label{ke}
\ee
and the numerical investigation has shown that the Newtonian filter
will determine the effective mass parameter which numerically value is
accurately given by the above analytic formula.
The $k_E$ parameter can be used to give an estimate of the chirp mass
of the binary system. We define {\em generalized chirp mass} ${\cal M}_g$ as
\begin{equation}
{\cal M}_g = 1/k_E^{3/5}
\end{equation}
We have calculated numerically the generalized chirp mass using the analytic
formula (\ref{ke}) and we have found
that it deviates from the true value by less than 4\%
for the range of masses from 1.4 to 10 solar mass.
For the range of masses from 1.01 to 1.64 which is the expected range
of neutron star masses given present observations of binary pulsars \cite{Fb}
the generalized chirp
mass is always less than the true one by around 4\% but
with a very small range of .5\% around the average value.

Because of the inequality $m \geq 2^{6/5}{\cal M}$
and the closeness of the generalized chirp
mass to the true chirp mass the generalized chirp mass
${\cal M}_g$ gives a lower bound on the total mass
of the system. Thus from its estimate we can determine what binary
system we observe.
Also the R.H.S. of the above inequality gives
a poor man's estimate of the total mass.
For the range of masses of (1$M_{\odot}$,10$M_{\odot}$)
it deviates by 50\% from the true value
of the total mass but for the range of
(1.01$M_{\odot}$, 1.64$M_{\odot}$) acceptable for
neutron star binaries it is only 5\% smaller than the true mass.

Another application of this estimate is that it can be used
as an additional check on whether we
are observing the real signal.
If our estimate would deviate unusually from the predicted
range of ${\cal M}_g$ corresponding to the range of
individual masses of (1$M_{\odot}$,10$M_{\odot}$)
we could veto the detection.

An interesting application of the Newtonian filter would be to determine
unexpected effects in the binary interaction
that we would not be able to model and
introduce into multiparameter numerical templates
because we do not know their form.
The idea is to use the estimates of the effective mass
parameter $k_E$. Particularly useful
would be estimates of $k_E$ in the case of neutron star binaries.
Since the range of the neutron
star masses in a binary system is rather narrow the range of the allowable
values for the generalized chirp mass
will also be narrow. From the analysis in \cite{Fb} the range
from the least lower bound and to the greatest upper bound is
(1.01$M_{\odot}$, 1.64$M_{\odot}$)
and the range from greatest lower bound to least upper bound is as narrow as
(1.34$M_{\odot}$, 1.43$M_{\odot}$). This implies
the respective ranges in $k_E$ to be (0.57, 1.26) and (0.79, 0.71).
 From the population of estimates of the parameter $k_E$
we can determine its probability distribution and also the mean, variance
or range of observed values of $k_E$.
One can then compare the observed distribution of
$k_E$ and its characteristics with the ones obtained from observations
of the neutron star binaries in
our Galaxy or from the theoretical analysis and search for differences.
As an example we consider Damour-Esposito-Far\'ese bi-scalar tensor theory
described at the end of Section 2.4. The shift in the Newtonian mass
parameter $k$ due
to effects of this theory is given by formula (\ref{sh}).
We have calculated this shift numerically and we have found that for
the range of neutron star masses (1.01$M_{\odot}$ , 1.64$M_{\odot}$)
and the parameter $\beta$ = 1 (current observational bound)
the shift is in the range of (0.018, 0.022). This shift is much larger
than rms error in estimation of $k_E$ of 0.00037 (see Table XI).
Consequently the effects of the bi-scalar theory could be
determined to an accuracy depending on how well we would know the
probability distribution of the neutron star masses
and the number of available detections of gravitational waves
from binaries.

\subsection{Post-Newtonian search templates}

In a recent work \cite{Apo} different search templates
than the Newtonian filter
were recommended and extensively analysed.
The proposed templates are the post-Newtonian waveforms
with all the spin effects and parameters removed. They have four parameters:
amplitude, phase, reduced mass, total mass.
We shall denote such search templates by 1PNf, 3/2PNf, 2PNf where the
number in front refers to the order of post-Newtonian effects included.
In Ref.\cite{Apo} the fitting factor FF ($FF = l^2$ see Appendix D)
of the 3/2PNf search template was
calculated and it was concluded that this template family works quite
well even for signals with with both spin-modulational and the nonmodulated
3/2 post-Newtonian effects combined.
In this Appendix we investigate the performance of the 2PNf search
template for the case of the 2nd post-Newtonian signal in the approximation
considered in Section 2. This means that we ignore all post-Newtonian
effects in the amplitudes of both the signal and the template and we
assume that the spin-orbit and the spin-spin parameters $s_o$ and $s_s$
in the signal are constant.
In Table XII  we give the factor $l$ and the shift in the time parameter,
phase, reduced mass and total mass for the three representative binary
systems described in Section 2. We have also given the shifts in the reduced
and the total mass parameters in percentages of their true values.

\bigskip

{\bf Table XII.}
{\em Performance of the 2nd post-Newtonian search template for
the three fiducial binaries located at the distance of 200Mpc.}

\bigskip

\begin{tabular}{|c|c|c|c|c|c|c|c|} \hline
Binary & l & $\delta\mu$ & $\frac{\delta\mu}{\mu}$ & $\delta m$ &
$\frac{\delta m}{m} $ & $\delta t$[ms] & $\delta\phi$ \\ \hline
NS-NS & 0.98 & 0.0028 & 0.5\% & -0.017 & 0.61\% & -9.5$\times 10^{-3}$
& 0.00027 \\
NS-BH & 0.95 & 0.52 & 42\% & -4.8 & 42\% & -3.0 & -0.28 \\
BH-BH & 0.98 & 1.9 & 38\% & -7.8 & 39\% & 2.3$\times 10^{-3}$
& -0.00053 \\ \hline
\end{tabular}

\bigskip

We see that the 2PNf search template fits the signal better than the
Newtonian search template Nf investigated in Section 3.1.
There are two reasons
for this. The 2PNf template has one more parameter than Nf template and
the phase of 2PNf template has all post-Newtonian frequency evolution terms
whereas the phase of the Nf template has only Newtonian frequency evolution
$f^{-5/3}$. Also in the case of NS-NS binary which has small spin parameters
the expectation values of the estimates
of the reduced and the total masses are close to their true values.

The advantage of the Newtonian search template might be its simplicity:
it has the least possible number of parameters
and hence the least computational time is needed to implement such a template
in data analysis algorithms.
Before the detailed data analysis schemes are developed for the real detectors
it is useful to investigate theoretically a wide range of possible
search templates.

We have also calculated the covariance matrix for the 2PNf template.
The results are summarized in Table XIII where we have given the rms errors
in the time, reduced mass and the total mass parameters of this search
template for the three binary systems. We have also given the errors in
the reduced and the total mass in percentage of their true values.

\bigskip

{\bf Table XIII.}
{\em The rms errors in the estimators of the parameters of
the 2nd post-Newtonian search template for the three fiducial binary systems
located at the distance of 200Mpc.}

\bigskip

\begin{tabular}{|c|c|c|c|c|c|} \hline
Binary & $\Delta t_a$[ms] & $\Delta\mu_{PN}$$[M_{\odot}]$ &
$\Delta m_{PN}$$[M_{\odot}]$ & $\frac{\Delta \mu_{PN}}{\mu}$
& $\frac{\Delta m_{PN}}{m}$ \\ \hline
NS-NS & 0.80 & 0.0078 & 1.1\% & 0.011 & 0.39\% \\
NS-BH & 0.40 & 0.012 & 1.0\% & 0.0068 & 0.06\% \\
BH-BH & 0.16 & 0.0090 & 0.2\% & 0.0050 & 0.03\% \\ \hline
\end{tabular}

\bigskip

We see that the rms errors of the parameters of the post-Newtonian search
template are comparable to rms errors obtained with optimal filtering of
the signal with spin parameters removed.

\bigskip

\section{Conclusions}

The analysis of the accuracy of estimation of parameters of the 2nd
post-Newtonian signal (Section 2.3) has shown that main characteristics of
this signal: chirp mass and the time parameter can be estimated to
a very good accuracy: chirp mass to 0.1\% - 1.0\% amd time parameter
to a quarter of a millisecond for typical binaries.
A typical binary consists of compact objects of 1.4 to 10 solar masses
and is located
at the distance of 200Mpc from Earth and the amplitude of its
gravitational wave signal is averaged over all directions and orientations.
The signal-to-noise ratio of typical binaries varies from 15 to 77 for
the planned advanced LIGO interferometers.
However the accuracy of determination of post-Newtonian effects
is considerably degraded due to large number parameters: 6 parameters
in the phase of the 2nd post-Newtonian signal (Table II).
Consequently the errors in determination of the reduced mass and the total
mass are large and range from 50\% to 200\% for typical systems (Table IV).
If spin effects could be neglected thereby
reducing the number of parameters by 2 the rms errors of estimation of
reduced and total masses would have a very impressive value
of a fraction of a percent (Table V).

Analysis of the accuracy of estimation of the effects of the dipole radiation
in the Jordan-Fiertz-Brans-Dicke theory of gravity has shown that
the planned laser interferometric gravitational wave detectors
should have ability of testing alternative theories of gravity comparable to
that of current observations in the solar system and our Galaxy.

The numerical analysis of Section 2 supports the need for the
search templates emphasized in Ref.\cite{Cal}. The results of Section
3 show that the Newtonian filter (a search template with only one mass
parameter) will perform reasonably well at least for the case of
of constant spin parameters. Such a filter can be used to perform an on
line scan of the data to search for the candidates for real signals.
The measurement of the mass parameter of the Newtonian signal
provides an accurate estimate of an effective mass parameter $k_E$
of the binary (see Eq.\ref{ke}). The value of this parameter gives
the information about the binary analogous to the chirp mass in the analysis
of the signal in the quadrapole approximation.
Moreover this parameter contains information
about the post-Newtonian effects and it can contain information
about the effects that we cannot at present model for example
about the effects due
to unknown corrections to general relativity in the strong field regime.
Such information can be extracted if we built a probability distribution
of $k_E$ from its estimators by the Newtonian filter.
The post-Newtonian search templates analysed in \cite{Apo} perform better
than Newtonian filters and considering increasing computational capability
they can also be used in the on line analysis of the data.
In the case of large spin
parameters it would be useful to obtain relations of the two mass
parameters in
such templates to the true masses and spins similar to relation of
the effective mass parameter of the Newtonian filter
to the other parameters of the binary (see Eq. \ref{ke}).
For the case of the observed
binary systems, binaries consisting of two neutron stars with small spin
parameters the Newtonian filter will provide an accurate
estimate of the chirp mass whereas the post-Newtonian search templates
will provide accurate estimates of reduced and total masses.

\bigskip

{\bf Acknowledgment}

\bigskip

One of us (A. Kr\'olak) would like to thank the Max-Planck-Gesellschaft
for support and the Arbeitsgruppe Gravitationstheorie an der Friedrich-
Schiller-Universit\"at in Jena for hospitality during the time this
work was done.
We would like to thank T. Apostolatos and K. S. Thorne for
helpful discussions.
This work was supported in part by KBN grant No. 2 P302 076 04.

\newpage

\newcommand{\N}{\mbox{\rm I\hspace{-0.2em}N}}
\newcommand{\R}{\mbox{\rm I\hspace{-0.2em}R}}
\newcommand{\C}{\mbox{\rm \hspace{0.31em}\rule{.5pt}{7pt}\hspace{-0.31em}C}}
\newcommand{\mfrac}[2]{\mbox{$\frac{#1}{#2}$}}
\newcommand{\Arccos}{\mbox{\rm Arccos}}
\newcommand{\Arctan}{\mbox{\rm Arctan}}
\newcommand{\Od}[1]{{\cal O}\left(#1\right)}
\newcommand{\bm}[1]{{\mbox{\boldmath$#1$}}}
\newcommand{\mbf}[1]{{\bf #1}}
\newcommand{\mcal}[1]{{\cal #1}}
\newcommand{\DS}{\displaystyle}
\newcommand{\mf}[1]{\mbox{\fbox{$\displaystyle #1$}}}

\renewcommand{\thefootnote}{\fnsymbol{footnote}}
\renewcommand{\d}{{\rm d}}
\bigskip

{\bf Appendix A: The effects of eccentricity}

\bigskip

In this appendix we derive the first order correction due to eccentricity
in the phase of the gravitational wave signal from a binary system.
The derivation is due to N. Wex \cite{NW}.

Let $a$ and $e$ be respectively the semi-major axis and the eccentricity of
the Keplerian
orbit of a binary. From the quadrupole formula one obtains the following
expressions for the secular changes of $a$ and $e$ averaged over an orbit
\cite{PM}
\begin{equation} \label{dandt}
   \left\langle\frac{\d a}{\d t}\right\rangle = -\frac{\beta}{a^3}\,
   \frac{1+\mfrac{73}{24}e^2+\mfrac{37}{96}e^4}{(1-e^2)^{7/2}}, \qquad
   \left\langle\frac{\d e}{\d t}\right\rangle = -\frac{304}{15}\,
  \frac{\beta}{a^4}\,\frac{e\left(1+\frac{121}{304}e^2\right)}{(1-e^2)^{5/2}}.
\end{equation}
where $\beta = \frac{64}{5}\,m^2\mu$. From these equations we get $\d
a/\d e$ which can be integrated with respect to $e$. The result is:
\begin{equation} \label{a(e)}
a(e) = a_0\,\frac{\xi(e)}{\xi(e_0)}, \qquad \xi(e) \equiv e^{12/19}
       \frac{\left(1+\frac{121}{304}e^2\right)^{870/2299}}{1-e^2}
\end{equation}
where $e_0$ is an arbitrary initial eccentricity
and $a_0 = a(e_0)$. From Kepler's
third law $\pi f=m^{1/2}a^{3/2}$, where $f$ is gravitational
wave frequency we get an analytic expression for $f$ as a function of $e$.
\begin{equation} \label{f(e)}
f(e) = f_0\,\frac{\eta(e_0)}{\eta(e)}, \qquad \eta(e) \equiv e^{18/19}
       \frac{\left(1+\frac{121}{304}e^2\right)^{1305/2299}}{(1-e^2)^{3/2}}
\end{equation}
where $f_0 = f(e_0)$.
For small eccentricities we find
\begin{equation}
e = e_0\left(\frac{f}{f_0}\right)^{-19/18}\left[1+\Od{e_0^2}\right].
\label{f(e)}
\end{equation}
Thus to first order in $e$ the quantity $I_e = e^2_0 f_0^{19/9}$ is a constant.
We call $I_e$ the {\em asymptotic eccentricity invariant}.
The {\em characteristic time} for the evolution of
the binary system is given by
\begin{equation}
\tau_e:=\frac{f}{\d f/\d t}=f\left(\frac{\d f}{\d a}\,
                            \frac{\d a}{\d t}\right)^{-1}.
\end{equation}
{}From Kepler's third law we find
\begin{equation}
 \tau_e = \frac{5}{96}\frac{1}{\mu m^{2/3}}\frac{1}{(\pi f)^{8/3}}\,
         \frac{(1-e^2)^{7/2}}{1+\mfrac{73}{24}e^2+\mfrac{37}{96}e^4}
\end{equation}
For small eccentricities $e$ we get
\begin{equation}
   \tau_e = \frac{5}{96}\frac{1}{\mu m^{2/3}}\frac{1}{(\pi f)^{8/3}}\,
   \left[1-\frac{157}{24}e^2+\Od{e^4}\right].
\end{equation}
Therefore using Eq.\ref{f(e)} we can express the characteristic time
with first order correction due to eccentricity as
\begin{equation}
\tau_e = \frac{5}{96}\frac{1}{\mu m^{2/3}}\frac{1}{(\pi f)^{8/3}}\,
   \left[1-\frac{157}{24}e_0^2\left(\frac{f}{f_0}\right)^{-19/9}\right].
\end{equation}
The phase of the Fourier transform of the signal in the stationary phase
approximation is given by
\begin{eqnarray}
&&\varphi[f] = 2\pi f t_i-\varphi_i- \pi/4
- 2\pi\int_{f_i}^f\tau_e(f')(1-f/f')\,\d f =
\nonumber\\&& = 2\pi f t_a - \varphi +
 \frac{1}{128\mu m^{2/3}} \times
 \left[\left(\frac{3}{(\pi f)^{5/3}}+\frac{5\pi f}{(\pi f_a)^{8/3}}
 -\frac{8}{(\pi f_a)^{5/3}}\right)\right. \nonumber\\&&
 -\left.\frac{785}{1462}e_0^2(\pi f_0)^{19/9}\left(\frac{9}{(\pi f)^{34/9}}
 +\frac{34\pi f}{(\pi f_a)^{43/9}}-\frac{43}{(\pi f_a)^{34/9}}\right)\right]
\end{eqnarray}
and consequently the Fourier transform of our signal
in the stationary phase approximation has the form
\begin{equation}
\tilde h(f) = \mcal{A} f^{-7/6}\exp i\left[2\pi f t_a-\varphi-\pi/4+
\frac{5}{48}(a(f;f_a)k+a_e(f;f_a)k_e)\right], \qquad \mbox{for} \quad f>0.
\end{equation}
(and by the complex conjugate of the above expression for $f<0$) where
\begin{eqnarray}
&& \tilde{A} =
\frac{1}{30^{1/2}}\frac{1}{\pi^{2/3}}\frac{\mu^{1/2}m^{1/3}}{R},\\
&& k = \frac{1}{\mu m^{2/3}} , \hspace{10mm}
k_e=\frac{1}{\mu m^{2/3}}e_0^2(\pi f_0)^{19/9}, \\
&& a(f;f_a) = \frac{9}{40}\frac{1}{(\pi f)^{5/3}} +
\frac{3}{8}\frac{\pi f}{(\pi f_a)^{8/3}}
- \frac{3}{5}\frac{1}{(\pi f_a)^{5/3}}, \\
&& a_e(f;f_a) = - \frac{157}{24}\left(\frac{81}{1462}\frac{1}{(\pi f)^{34/9}}
+ \frac{9}{43}\frac{\pi f}{(\pi f_a)^{43/9}} -
\frac{9}{34}\frac{1}{(\pi f_a)^{34/9}}\right).
\end{eqnarray}

We have investigated the accuracy of measurements of parameters of
the above signal with first order eccentricity contribution.
We have considered neutron star/neutron star binary. The results are
summarized in Table XIV.

\bigskip

{\bf Table XIV}
{\em The rms errors of the parameters of the signal with first order
contribution due to eccentricity for a binary of two neutron stars of
1.4 solar mass each at the distance of 200Mpc.}

\begin{center}
\begin{tabular}{|c|c|c|c|c|c|} \hline
S/N & $\Delta t_c$[ms] & $\Delta\phi_c$
& $\Delta \mu/\mu$ & $\Delta m/m$
& $\Delta k_e [M^{-5/3}_{\odot}$(100Hz)$^{19/9}]$ \\ \hline
15 & 0.56 & 1.2 & 0.50\% & 0.74\% & $3.6 \times 10^{-7}$ \\ \hline
\end{tabular}
\end{center}

However for the currently observed binaries the eccentricity invariant
$I_e$ is extremely small. For Hulse-Taylor pulsar
$I_e = 1.8 \times 10^{-13} [M^{-5/3}_{\odot}$100Hz$^{19/9}$].
We have the following numerical values.
\bea
&& k_e = 1.3\times 10^{-13}(\frac{I_e}{1.8\times 10^{-13}})
(\frac{1.2}{{\cal M}_{\odot}})^{5/3}[M^{-5/3}_{\odot}\mbox{100Hz}^{19/9}] \\
&& \frac{\Delta k_e}{k_e} =
2.8\times 10^6 r_{200Mpc}(\frac{I_e}{1.8\times 10^{-13}})
(\frac{{\cal M}_{\odot}}{1.2})^{5/3},
\eea
where $r_{200Mpc}$ is distance in 200Mpc.
Thus for eccentricity effects to be measured one would need extremely
short period binaries of high eccentricity. Such binaries could perhaps
occur in the center of a galaxy or be created as a result of some
supernova explosions.

\bigskip

{\bf Appendix B: Covariance matrices at various post-Newtonian orders}

\bigskip

In this Appendix we give the numerical values of the covariance matrices
at various post-Newtonian orders for the reference binary.
The reference binary has the chirp mass ${\cal M}$ of 1 solar mass and is
located at the distance of 100Mpc. We only give reduced covariance matrices
i.e. covariance matrices for the phase parameters. As indicated in Section 2
the estimator of the amplitude parameter is uncorrelated with
phase parameters.
The integration range in the Fisher matrix integrals was taken to be from 10Hz
to infinity and the spectral density of advance LIGO detectors was assumed
(Eq.\ref{Sh}). The frequency $f_a$ was chosen such that the rms error
in the time parameter is minimum. The minimum frequency is denoted by $f_m$
and its numerical value is given for each covariance matrix.
The subscripts N, 1PN, 3/2PN, 2PN, 2PNe refer to signal including
quadrupole radiation, 1st post-Newtonian correction, 3/2 post-Newtonian
correction, 2nd post-Newtonian correction,
and 1st order effect due to eccentricity respectively.
The order of parameters in the matrices is the following:
$t_c$, $\phi_c$, $k$, $k_1$, $k_{3/2}$, $k_2$, $k_e$.

$f_m^N = 70$Hz

\be
C_N =  \left(\begin{array}{ccc}
 1.96\times 10^{-8} & 8.2\times 10^{-6} & 1.21\times 10^{-10} \\
 8.2\times 10^{-6} & 0.00537 & 2.08\times 10^{-7} \\
 1.21\times 10^{-10} & 2.08\times 10^{-7} & 6.65\times 10^{-11}
\end{array}\right)
\ee

$f_m^{1PN} = 100$Hz

\be
C_{1PN} = \\
\left(\begin{array}{cccc}
2.24\times 10^{-8} & 9.93\times 10^{-6} &
8.87\times 10^{-10} & -6.61\times 10^{-8}  \\
9.93\times 10^{-6} & 0.00754 & -6.28\times 10^{-7} & 0.000147 \\
8.87\times 10^{-10} & -6.28\times 10^{-7} & 1.4\times 10^{-9} &
-1.97\times 10^{-7} \\
-6.61\times 10^{-8} & 0.000147 & -1.97\times 10^{-7}
& 0.0000289
\end{array}\right)
\ee

$f_m^{3/2PN} = 160$Hz
{\small
\be
C_{3/2PN} = \\
\left(\begin{array}{ccccc}
3.41\times 10^{-8} & 0.00002 &
-9.65\times 10^{-9} & 3.36\times 10^{-6} &  0.0000212 \\
0.00002 & 0.0191 & -9.34\times 10^{-7} & -0.000403 & -0.007 \\
-9.65\times 10^{-9} & -9.34\times 10^{-7} &
2.23\times 10^{-8} & -8.71\times 10^{-6} & -0.0000622 \\
3.36\times 10^{-6} & -0.000403 & -8.71\times 10^{-6} & 0.00349
& 0.0253  \\
0.0000212 & -0.007 & -0.0000622 & 0.0253 & 0.185
\end{array}\right)
\ee}

$f_m^{2PN} = 100$Hz
{\small
\be
C_{2PN} = \\
\left(\begin{array}{cccccc}
5.62\times 10^{-8} & 0.0000286 &
2.39\times 10^{-9} & -9.48\times 10^{-6} & -0.000199 & -0.00102 \\
0.0000286 & 0.0189 & 0.000016 & -0.0162 & -0.26 & -1.11  \\
2.39\times 10^{-9} & 0.000016 & 2.3\times 10^{-7} & -0.000162
& -0.00219 & -0.00796 \\
-9.48\times 10^{-6} & -0.0162 & -0.000162 & 0.116 & 1.59 & 5.86 \\
-0.000199 & -0.26 & -0.00219 & 1.59 & 22. & 81.5  \\
-0.00102 & -1.11 & -0.00796 & 5.86 & 81.5 & 305.
\end{array}\right)
\ee}


$f_m^{2PNe} = 120$Hz

$C_{2PNe}$ =
{\small
\be
\left(\begin{array}{ccccccc}
6.36\times 10^{-8} & 0.0000369 &
-3.56\times 10^{-8} & 4.9\times 10^{-7} & -0.000135 & -0.000977 &
-4.86\times 10^{-11} \\
0.0000369 & 0.0279 & 0.000035 & -0.0315 & -0.474 & -1.89
& -5.21\times 10^{-9} \\
-3.56\times 10^{-8} & 0.000035 & 2.36\times 10^{-6} & -0.00121 &
-0.0143 & -0.0464 &  1.33\times 10^{-9} \\
4.9\times 10^{-7} & -0.0315 & -0.00121 & 0.63 & 7.57 & 24.7
& -6.52\times 10^{-7} \\
-0.000135 & -0.474 & -0.0143 & 7.57 & 91.5 & 301.
& -7.59\times 10^{-6} \\
-0.000977 & -1.89 & -0.0464 & 24.7 & 301. & 999. & -0.000024 \\
-4.86\times 10^{-11} & -5.21\times 10^{-9} & 1.33\times 10^{-9} &
-6.52\times 10^{-7} & -7.59\times 10^{-6} & -0.000024
& 8.28\times 10^{-13}
\end{array}\right)
\ee}

\bigskip

\bigskip

{\bf Appendix C : Coefficients in the Damour-Esposito-Far\'ese
biscalar T($\beta',\beta''$) theory}

\bigskip

The coefficients $\kappa_o, \kappa_q, \kappa_{d1}, \kappa_{d2}$
in the shift of the Newtonian mass parameter $k$
due to the biscalar T($\beta',\beta''$) theory (Eq.\ref{sh} in Section 3.4)
are given by the following formulae
\bea
\kappa_o &=& \frac{1}{2} \beta' B ({\cal C}_1^2 + {\cal C}_2^2), \\
\kappa_q &=& \beta' B ({\cal C}_1^2 x_2 + {\cal C}_2^2 x_1), \\
\kappa_{d1} &=& \frac{1}{2} \beta' B ({\cal C}_1^2 x_1 - {\cal C}_2^2 x_2)
(x_1 - x_2), \\
\kappa_{d2} &=& (ab_{121} - ab_{221}) x_1 + (ab_{212} - ab_{112}) x_2,
\eea
where
\bea
x_1 = \frac{m_1}{m}, \\
x_2 = \frac{m_2}{m},
\eea
and constant A and B have the values
\be
A = 2.1569176, \hspace{10mm}
B = 1.0261529.
\ee
${\cal C}_1$ and ${\cal C}_2$ are sensitivities of the two bodies to changes
of the scalar field.
The functions $ab$ are given by
\hspace*{-15mm}
\bea
ab_{121} &=& \beta' (-{\cal C}_2 - B {\cal C}_1^2 + (A - 3 B) {\cal C}_2^2
- (A - B) 2 {\cal C}_2 {\cal C}_1^2 + \\ \nonumber
& &(2 A^2 - 7 A B + 5 B^2) {\cal C}_2^2 {\cal C}_1^2 ) +
{\beta'}^2 B^2 ( - 3 {\cal C}_1^3 + 2 {\cal C}_1^2 {\cal C}_2^2 +
{\cal C}_1^4 + \frac{1}{2} {\cal C}_2 {\cal C}_1^4 +
A {\cal C}_2^2 {\cal C}_1^4) +
\frac{1}{2} \beta'' B {\cal C}_2^2, \\
ab_{212} &=& \beta'(-{\cal C}_1 - B {\cal C}_2^2 + (A - 3 B) {\cal C}_1^2
- (A - B) 2 {\cal C}_1 {\cal C}_2^2 + \\ \nonumber
& &(2 A^2 - 7 A B + 5 B^2) {\cal C}_1^2 {\cal C}_2^2 ) +
{\beta'}^2 B^2 ( - 3 {\cal C}_1^3 + 2 {\cal C}_1^2 {\cal C}_2^2
+ {\cal C}_1^4 + \frac{1}{2} {\cal C}_1 {\cal C}_2^4 +
A {\cal C}_1^2 {\cal C}_2^4) + \frac{1}{2} \beta'' B {\cal C}_1^2 , \\
ab_{221} &=& \beta'(-{\cal C}_2 - \frac{1}{2} B ({\cal C}_1^2
+ {\cal C}_2^2) + (A - 3 B) {\cal C}_2^2 -
(A - B) {\cal C}_2 ({\cal C}_1^2 + {\cal C}_2^2) + , \\ \nonumber
& &\frac{1}{2}(2 A^2 - 7 A B + 5 B^2) {\cal C}_2^2 ({\cal C}_1^2
+ {\cal C}_2^2)) +
{\beta'}^2 B^2 ( - 3 {\cal C}_2^3 + {\cal C}_2^2 ({\cal C}_1^2
+ {\cal C}_2^2) +, \\ \nonumber
& & {\cal C}_2^4 + \frac{1}{2} {\cal C}_2^3 {\cal C}_1^2 +
A {\cal C}_1^2 {\cal C}_2^4) + \frac{1}{2} \beta'' B {\cal C}_2^2, \\
ab_{112} &=& \beta'(-{\cal C}_2 - \frac{1}{2} B ({\cal C}_1^2
+ {\cal C}_2^2) + (A - 3 B) {\cal C}_1^2 -
(A - B) {\cal C}_1 ({\cal C}_1^2 + {\cal C}_2^2) + , \\ \nonumber
& &\frac{1}{2}(2 A^2 - 7 A B + 5 B^2) {\cal C}_1^2 ({\cal C}_1^2
+ {\cal C}_2^2)) +
{\beta'}^2 B^2 ( - 3 {\cal C}_1^3 + {\cal C}_1^2 ({\cal C}_1^2
+ {\cal C}_2^2) + , \\ \nonumber
& &{\cal C}_1^4 + \frac{1}{2} {\cal C}_2^2 {\cal C}_1^3 +
A {\cal C}_2^2 {\cal C}_1^4) + \frac{1}{2} \beta'' B {\cal C}_1^2 .
\eea
For a detailed exposition of the theory the reader
should consult Ref.\cite{DEF}.

\bigskip

{\bf Appendix D: Detection of the known signal with a non-optimal filter}

\bigskip

Suppose that we would like to know whether or not in a given
data set $x$ there is present a signal $h$.
We assume that the noise $n$ in the data is additive.
There are two alternatives:
\bea
\mbox{NO SIGNAL} &:&  x = n  \nonumber \\
\mbox{SIGNAL}    &:&  x = h + n
\eea
A standard method to determine which of the two alternatives holds
is to perform the Neyman-Pearson test{\cite{D}}. This test consists
in comparing the {\em likelihood ratio} $\Lambda$, the ratio of probability
density distributions of the data $x$
when the signal is present and when the signal is absent,
with a threshold. The threshold is determined by the false alarm probability
that we can tolerate (the false alarm probability is the probability of
saying that the signal is present when there is no signal).
The test is optimal in the sense that it maximizes
the probability of detection of the signal.
In the case of Gaussian noise and deterministic signal
$h$ the logarithm of $\Lambda$ is given by
\be
\ln\Lambda = (x|h) - \frac{1}{2}(h|h).
\ee
Thus in this case the optimal test consists of correlating the data
with the expected signal and it is equivalent to comparing the correlation
$G := (x|h)$ with a threshold.
The probability distributions $p_0$ and $p_1$ of $G$ when respectively
the signal is absent and present are given by
\bea
p_0(G;d) &=& \frac{1}{\sqrt{2\pi d^2}}
\exp\left[-\frac{G^2}{d^2}\right], \\
p_1(G;d) &=& \frac{1}{\sqrt{2\pi d^2}}
\exp\left[-\frac{(G - d^2)^2}{d^2}\right],
\eea
where d is the optimal signal-to-noise ratio $d^2 = (h|h)$ and we assumed that
the noise is a zero mean Gaussian process.

Let $T$ be a given threshold. This means that we say that the signal is
present
in a given data set if $G > T$. The probabilities $P_F$ and $P_D$ of false
alarm and detection respectively are given by
\bea
P_F(T,d) &=& \int_{T}^{\infty} p_0(G;d)\, dG, \\
P_D(T,d) &=& \int_{T}^{\infty} p_1(G;d)\, dG
\eea
In the Gaussian case they can be expressed in terms of the error functions.
\bea
P_F(d_T,d) = \frac{1}{2} erfc(\frac{d_T^2}{\sqrt{2} d}), \label{pf} \\
P_D(d_T,d) = \frac{1}{2} (1 + erf(\frac{d^2 - d_T^2}{\sqrt{2} d}),
\eea
where erf and erfc are error and complementary error functions
respectively \cite{err}.
and we have introduced for convenience the quantity $d_T := \sqrt{T}$
that we call the {\em threshold signal-to-noise ratio}.
In practice we adopt a certain value of the false alarm probability
that we can accept
and from formula (\ref{pf}) we calculate the detection threshold T.

Let $F$ be a linear filter and let $n$ be the additive noise in data $x$ then
\be
(x|F) = (s|F) + (n|F).
\ee
The signal-to-noise (S/N) ratio is defined by
\be
(S/N)^2 := \frac{E_1[(s|F)^2]}{E_1[(n|F)^2]} = \frac{(s|F)^2}{(F|F)}.
\ee
where $E_1$ means expectation value when the signal is present.
By Schwartz inequality we immediately see that (S/N) is maximal and
equal to $d$ when the linear filter is matched to the signal
i.e. $F = h$.
This is another interpretation of the matched filter - it maximizes
the signal-to-noise ratio over all linear filters \cite{D}.
However when the noise is not Gaussian the matched filter is
not the optimal filter;
it does not maximize probability of detection of the signal.
We see that in the case of Gaussian noise the problem of detecting
a known signal by optimal filter is determined by one parameter -
the optimal signal-to-noise ratio $d$.

Suppose that because of certain restrictions of practical nature
we cannot afford to use the optimal filter $h$ and we use
a suboptimal one - $h_N$ which is not perfectly matched to the signal.
Thus $(h|h_N) < (h|h)$. We denote $\sqrt{(h|h_N)}$ by $d_o$
and we assume that $(h_N|h_N) = (h|h) = d^2$.
Our suboptimal correlation function is given by $G_N = (x|h_N)$
and its probability distributions $p_{N0}$ and $p_{N1}$
when respectively the signal is absent and present are given by
\bea
p_{N0}(G_N;d) &=& \frac{1}{\sqrt{2\pi d^2}}
\exp\left[-\frac{G_N^2}{d^2}\right], \\
p_{N1}(G_N;d,d_o) &=& \frac{1}{\sqrt{2\pi d^2}}
\exp\left[-\frac{(G_N - d_o^2)^2}{d^2}\right].
\eea
We see that the suboptimal detection problem is determined by two parameters
- $d$ and $d_o$, square roots of the expectation values of
the optimal and suboptimal correlations when the signal is present.
The false alarm and detection probabilities as in the optimal case can be
expressed in terms of the error functions.
\bea
P_F(d_T,d) = \frac{1}{2} erfc(\frac{d_T^2}{\sqrt{2} d}), \\
P_D(d_T,d,d_o) = \frac{1}{2} (1 + erf(\frac{d_o^2 - d_T^2}{\sqrt{2} d}).
\eea
We see that the probability of false alarm for the suboptimal case is
the same as in the optimal case however the probability of detection in the
suboptimal case is always less then the probability of detection in
the optimal case since $d_o < d$ and the error function $erf(x)$
is an increasing function of the argument $x$.
The signal-to-noise ratio in the case of suboptimal linear filter $h_N$
is given by
\be
(S/N)^2 = \frac{(h|h_N)^2}{(h_N|h_N)} = d^2 (\frac{d_o}{d})^4.
\ee
Let us denote the ratio $d_o/d$ by $l$. The ratio $l$ measures the drop in the
expectation value of the correlation function as a result of non-optimal
filtering. We see that due to
suboptimal filtering the signal-to-noise ratio decreases by {\bf square}
of the factor $l$. We denote $l^2$ by FF and following Ref.\cite{Apo}
call it the {\em fitting factor}.

In our considerations we need to calculate the number of events that
will be detected by linear filtering.
We shall make a number of simplifying assumptions. We shall assume
a Euclidean universe where in the sphere of radius $r_o$ we have one source
and that at the distance $r_o$ the optimal signal-to-noise ratio is d.
Moreover we shall assume that the magnitudes of the signal $h$
and the suboptimal filter $h_N$ are inversely proportional to the distance
$r$ from the source.
Then the square roots $d_r$ and $d_{or}$ of the expectation
values of the optimal and suboptimal correlations
at the distance $r$ are given by
\bea
d_r = \frac{r_o}{r} d, \\
d_{or} = \frac{r_o}{r} d_o.
\eea
We assume that the sources are uniformly distributed in space.
Then the expected number of detected real events $N$ and $N_N$ in the optimal
and the suboptimal case respectively is given by
\bea
N(d_T,d) &=& \frac{4\pi \int_{0}^{\infty} r^2 P_D(d_T,d_r)\, dr}
{\frac{4\pi}{3}r_o^3} = \nonumber \\
& &3 \int_0^{\infty} x^2 P_D(d_T,d/x)\, dx \label{n} \\
N_N(d_T,d,d_o) &=&
\frac{4\pi \int_{r_o}^{\infty} r^2 P_{ND}(d_T,d,d_o/r)\, dr}
{\frac{4\pi}{3}r_o^3} = \nonumber \\
& &3 \int_0^{\infty} x^2 P_{ND}(d_T,d,d_o/x)\, dr .
\label{nn}
\eea
The assumptions that led to the above formulae mean that we neglect
general relativistic, cosmological and evolutionary effects.
Because of the noise even if there is no signal there is always a non zero
probability that the correlation function crosses the threshold. Thus there
will be a certain number $N_F$ of false events.
For a given optimal signal-to-noise
ratio and a threshold $d_T$ this number is the same for both the
optimal and suboptimal filter and it is given by
\bea
N_F = \frac{4\pi \int_{r_o}^{\infty} r^2 P_F(d_T,d/r)\, dr}
{\frac{4\pi}{3}r_o^3} = \\ \nonumber
3 \int_o^{\infty} x^2 P_F(d_T,d/x)\, dx.
\label{nf}
\eea
We observe that in the Gaussian case the integrals in the formulae (\ref{n}),
(\ref{nn}), and (\ref{nf}) are convergent even though
we integrate over the all infinite Euclidean volume.

\bigskip

{\bf Appendix E: An approximate formula for the correlation function.}

\bigskip

In this Appendix we shall derive an approximate formula for the
correlation integral.
Let us consider the expression for the correlation
function given by (\ref{cor}).
The integrand of the correlation integral is the product of the integrand of
the signal-to-noise integral $Ind(f)$ considered in Section 2
and oscillating factor.
We know that the $Ind(f)$ is a fairly sharply peaked around a certain
frequency $f'_o$
consequently to obtain a reasonable approximation we expand the phase
around the frequency $f'_o$.
Keeping only the terms to the second order we get
\begin{eqnarray}
\Phi(f) \simeq 2\pi (f - f'_o) \Delta t' - \Delta\phi'' + ,\\ \nonumber
\frac{5}{96}\frac{(f/f'_o - 1)^2}{(\pi f'_o)^{5/3}}\Delta k_{TB} +
+ \ O[(f/f'_o - 1)^3].
\end{eqnarray}
where
\be
\phi''_o = \phi_o - 2\pi f'_o t'_o
\ee
and
\begin{equation}
k_{E} = k
- \frac{157}{24}\frac{k_e}{(\pi f'_o)^{19/9}} + k_1(\pi f'_o)^{2/3}
- k_{3/2}(\pi f'_o) + k_2(\pi f'_o)^{4/3}
- \frac{5}{192}\frac{k_D}{(\pi f'_o)^{2/3}}.
\end{equation}
We shall call $k_{E}$ an {\em effective mass parameter}.
$\Delta k_{E}$ is the difference in the effective mass parameter
of the signal and the filter.
Thus in the above approximation the post-Newtonian signal can be
parametrized by one effective mass parameter $k_E$.
In other words the dimension of the parameter space
of the filters is effectively reduced.
This last interpretation has been emphasized
in \cite{S} where 1st post-Newtonian corrections to the phase were considered.
The mass parameter estimated by Newtonian filter
considered in Section 3 is just
the effective mass parameter. We stress that the parameter $k_{E}$
depends not only on the parameters of the two-body system but also
on the characteristic frequency $f'_o$ of the noise in the detector.

The next step is to obtain a manageable approximation to the
function $Ind(f)$.
We approximate it by a Gaussian function with
the mean equal to the frequency $f'_o$ and the standard
deviation equal to the HWHM $\sigma'_o$ of the function $Ind(f)$.
We extend the range of integration from $-\infty$ to $+\infty$.
We introduce a normalization factor
such that the integral of the approximate integrand is equal to the
optimal signal-to-noise ratio $d$.
It is then useful to introduce a reduced correlation integral
$H' = H/d^2$ where d is the S/N ratio.
Thus our approximate formula for the reduced correlation integral
takes the form
\be
H'_a = \frac{1}{\sqrt{2\pi\sigma^2}}\int^{+\infty}_{-\infty}
\exp[-(f - f'_o)/(2{\sigma'_o}^2)]
\cos[2\pi (f - f'_o) \Delta t' + \Delta\phi'' +
\frac{5}{96}\frac{(f/f'_o - 1)^2}{(\pi f'_o)^{5/3}}\Delta k_{TB}]
\ee
The above integral can be done analytically. It is convenient
to introduce the following new variables and new parameters
\begin{eqnarray}
y &=& \frac{f - f'_o}{\sqrt{2{\sigma'_o}^2}}, \\
\vartheta'' &=& \Delta\phi'', \\
\tau &=& 2 \pi \Delta t'\sqrt{2{\sigma'_o}^2}, \\
\kappa &=& \frac{5}{96}\frac{\Delta k_{E}}{{f'_o}^2
(\pi f'_o)^{5/3}}2{\sigma'_o}^2
\end{eqnarray}
then our integral takes a simple form
\begin{equation}
H'_a(\vartheta,\tau,\kappa)
= \frac{1}{\sqrt{\pi}}\int^{+\infty}_{-\infty}\exp[-y^2]
\cos[-\vartheta'' + \tau y + \kappa y^2] dy.
\end{equation}
We see that in the new variables introduced above
the reduced correlation integral is
independent of the characteristics of
the integrand $Ind(f)$ i.e. $f'_o$ and $\sigma'_o$.
The analytic formula for the function
$H'_a(\vartheta'',\tau,\kappa)$ is given by
\begin{equation}
H'_a(\vartheta'',\tau,\kappa)
= \frac{1}{(1 + \kappa^2)^{1/4}}\exp[-\frac{\tau^2}{4 (1 + \kappa^2)}]
\cos[1/2(\arctan\kappa - \frac{\kappa\tau^2}{2 (1 + \kappa^2)}) - \vartheta'']
\end{equation}
By appropriate transformations given in Section 2 we can obtain
approximate formulae to the correlation integral
for an arbitrary choice of the time and the phase parameters.
Let us first consider the transformation given by Eq.(\ref{ph}).
In the coordinates introduced above it takes the form
\be
\vartheta'' = \vartheta - \frac{1}{\sqrt{2}} \tau\rho
\ee
Then the approximate formula for the correlation function is given by
\begin{equation}
H'_a(\vartheta,\tau,\kappa)
= \frac{1}{(1 + \kappa^2)^{1/4}}\exp[-\frac{\tau^2}{4 (1 + \kappa^2)}]
\cos[1/2(\arctan\kappa - \frac{\kappa\tau^2}{2 (1 + \kappa^2)})
+ \frac{1}{\sqrt{2}} \tau\rho - \vartheta]
\end{equation}
We see that in these new coordinates for $\kappa=0$
the correlation function oscillates
with the maxima at the discrete values of $\tau$ coordinate given by
\be
\tau_{max} = \frac{2\sqrt{2}\pi}{\rho} n,
\label{tau}
\ee
where
\begin{eqnarray}
\rho = f'_o/\sigma_o
\end{eqnarray}
and $n$ is an integer. In the original coordinates Eq.(\ref{tau})
takes the form $\Delta t = 1/f'_o$.
Thus the correlation integral oscillates with the period
determined by the characteristic frequency of the noise of the detector
$f'_o$.

The expressions for the correlation function for
different choice of the time and the phase parameters
can be obtained by the following transformations.
These are transformations given
by Eqs.(\ref{tp1}) and expressed in our dimensionless coordinates.
\begin{eqnarray}
\vartheta = \vartheta' + \frac{3}{5}\kappa'\rho^2(1 - \delta^{5/3}) , \\
\tau = \tau' + \frac{3}{4\sqrt{2}} \kappa'\rho'(1 -\delta^{8/3})
\end{eqnarray}
where
\begin{eqnarray}
\delta = f'_o/f_a.
\end{eqnarray}
{}From the approximate formula for the correlation function obtained above
we see that
the correlation is given by the product of an oscillating
cosine function and an envelope.
In the cosine function there are oscillations with the period of  $1/f'_o$.
The envelope function is exponentially damped
if we move away from the maximum at
the center except for the direction given by $\tau = 0$ along which
the damping is least. The equation of the ridge $\tau = 0$ in the primed
coordinates is given by
\bea
\tau' = - \frac{3}{4\sqrt{2}} \kappa'\rho'(1 -\delta^{8/3}),
\end{eqnarray}
and in the original coordinates it takes the form
\be
\Delta t = \frac{5}{256} \frac{1}{(\pi f'_o)^{8/3}}
(1 - (\frac{f_a}{f'_o})^{8/3}) \Delta k_E.
\label{tt'}
\ee
Consequently we conclude that the general appearance of
the correlation function in
coordinates $\Delta t'$ and $\Delta k$ is a series of peaks
aligned along a straight line
given by Eq.(\ref{tt'}) above and occurring
with the period $1/f'_o$ in the time coordinate.
Numerical investigation shows that the correlation integral exhibits
these properties and that our analytic formula reproduces qualitatively
its behaviour.
The approximate formula obtained above may be a useful tool for
developing algorithms to recognize the chirp signal in a noisy data set.


\begin{thebibliography}{99}
\bibitem{err}
The error function $erf(x)$ is defined by
\be
erf(x) = \frac{2}{\sqrt{\pi}}\int^x_0\exp(-t^2)\,dt
\ee
and the complementary error function $erfc(x)$ is defined by
\be
erfc(x) = 1 - erf(x).
\ee
\end{thebibliography}
\end{document}